\documentclass[8.5pt,twoside,twocolumn]{article}
\usepackage[super,sort&compress,comma]{natbib} 
\usepackage[version=3]{mhchem}
\usepackage{times,mathptmx}

\usepackage{sectsty}
\usepackage{balance} 

\usepackage{graphicx} 
\usepackage{lastpage}
\usepackage[format=plain,justification=raggedright,singlelinecheck=false,font=small,labelfont=bf,labelsep=space]{caption}
\usepackage{fancyhdr}
\usepackage{amsfonts}
\usepackage{amsmath}
\usepackage{amssymb}
\renewcommand{\vec}[1]{{\bf{#1}}}
\def\l{\left}
\def\r{\right}

\newcommand{\onlinecite}[1]{\hspace{-1 ex} \nocite{#1}\citenum{#1}} 
\usepackage{subfigure}
\usepackage{multirow}
\usepackage{array}

\begin{document}

\thispagestyle{plain}
\fancypagestyle{plain}{
\renewcommand{\headrulewidth}{1pt}}
\renewcommand{\thefootnote}{\fnsymbol{footnote}}
\renewcommand\footnoterule{\vspace*{1pt}%
\hrule width 3.4in height 0.4pt \vspace*{5pt}} 
\setcounter{secnumdepth}{5}

\makeatletter 
\def\subsubsection{\@startsection{subsubsection}{3}{10pt}{-1.25ex plus -1ex minus -.1ex}{0ex plus 0ex}{\normalsize\bf}}  
\def\paragraph{\@startsection{paragraph}{4}{10pt}{-1.25ex plus -1ex minus -.1ex}{0ex plus 0ex}{\normalsize\textit}} 
\renewcommand\@biblabel[1]{#1}            
\renewcommand\@makefntext[1]%
{\noindent\makebox[0pt][r]{\@thefnmark\,}#1}
\makeatother 
\renewcommand{\figurename}{\small{Fig.}~}
\sectionfont{\large}
\subsectionfont{\normalsize} 

\renewcommand{\headrulewidth}{1pt} 
\renewcommand{\footrulewidth}{1pt}
\setlength{\arrayrulewidth}{1pt}
\setlength{\columnsep}{6.5mm}
\setlength\bibsep{1pt}

\twocolumn[
  \begin{@twocolumnfalse}
\noindent\LARGE{\textbf{Phase Transformations in Binary Colloidal Monolayers$^\dag$}}
\vspace{0.6cm}

\noindent\large{\textbf{
Ye Yang,$^{\ast}$\textit{$^{a}$} 
Lin Fu,\textit{$^{b\ddag}$} 
Catherine Marcoux,\textit{$^{c\ddag}$} 
Joshua E.~S.~Socolar,\textit{$^{c}$} 
Patrick Charbonneau,\textit{$^{b,c}$} 
and
Benjamin B.~Yellen\textit{$^{a,d,e}$}}}\vspace{0.5cm}

\vspace{0.6cm}

\noindent \normalsize{Phase transformations can be difficult to characterize at the microscopic level due to the inability to directly observe individual atomic motions.  Model colloidal systems, by contrast, permit the direct observation of individual particle dynamics and of collective rearrangements, which allows for real-space characterization of phase transitions.  Here, we study a quasi-two-dimensional, binary colloidal alloy that exhibits liquid-solid and solid-solid phase transitions, focusing on the kinetics of a diffusionless transformation between two crystal phases.  Experiments are conducted on a monolayer of magnetic and nonmagnetic spheres suspended in a thin layer of ferrofluid and exposed to a tunable magnetic field. A theoretical model of hard spheres with point dipoles at their centers is used to guide the choice of experimental parameters and characterize the underlying materials physics. When the applied field is normal to the fluid layer, a checkerboard crystal forms; when the angle between the field and the normal is sufficiently large, a striped crystal assembles. As the field is slowly tilted away from the normal, we find that the transformation pathway between the two phases depends strongly on crystal orientation, field strength, and degree of confinement of the monolayer. In some cases, the pathway occurs by smooth magnetostrictive shear, while in others it involves the sudden formation of martensitic plates. 
}
\vspace{0.5cm}
 \end{@twocolumnfalse}
  ]

\section{Introduction}
\footnotetext{\dag~Electronic Supplementary Information (ESI) available: movies of experiments and simulations of martensitic transformations. See Soft Matter, 2015, DOI: 10.1039/C5SM00009B}

\footnotetext{\textit{$^{a}$~Duke University, Department of Mechanical Engineering and Materials Science, Box 90300 Hudson Hall, Durham, NC 27708}}
\footnotetext{\textit{$^{b}$~Duke University, Department of Chemistry, Durham, NC 27708. }}
\footnotetext{\textit{$^{c}$~Duke University, Department of Physics, Durham, NC 27708. }}
\footnotetext{\textit{$^{d}$~Duke University, Department of Biomedical Engineering, Durham, NC 27708. }}
\footnotetext{\textit{$^{e}$~University of Michigan - Shanghai Jiao Tong University, Joint Institute, Shanghai Jiao Tong University, Shanghai, China. }}


Diffusionless transformations are a class of solid-solid phase transitions in which the crystal unit cell changes shape and internal structure, while keeping its stoichiometry constant. These include magnetostriction, where a crystal undergoes continuous shear, as well as martensitic transformations, which involve more complex particle rearrangements. Because these transformations do not require long-range diffusion, they are fast and repeatable, and thus have been exploited in a number of engineering applications, including actuation in shape-memory alloys,\cite{kainuma, tanaka} hardening in steel,\cite{angel} and heat transfer in giant magnetocaloric materials.\cite{moya, liu}  Some martensitic transitions are irreversible, as in the quench hardening of steel,\cite{angel} while others are reversible, as in shape memory alloys\cite{kainuma, tanaka, song} and giant magnetocaloric materials.\cite{moya, liu}  The ability to optimize different aspects of these transformations, such as the response intensity and the repeatability over many cycles, requires detailed knowledge of how the crystal structure changes due to temperature and mechanical, electrical, or magnetic stresses.\cite{bhattacharya}  Because of their important technological applications, martensitic transformations have been extensively studied. Their mesoscopic features have been probed by calorimetry,\cite{chernenko} acoustic emission,\cite{meyers} and in situ microscopy;\cite{chernenko, waitz, meyers} their microscopic dynamics have also been studied through numerical simulation of simple models.\cite{kadau} However, no experimental technique has yet accessed the transformation dynamics with atomic resolution.

Though significantly larger than atoms, colloidal particles represent an accessible model for studying phase transitions in condensed matter.  Being individually resolvable by optical microscopy, yet sufficiently small to form thermally equilibrated phases at room temperature, micron scale colloidal particles have yielded important insights into mechanical crystal growth,\cite{gasser, leunissen, chen_Q, evers, talapin, shevchenko, shereda} melting in confined geometries,\cite{bubeck, wang} and solid-solid transitions.\cite{yethiraj_wouterse, yethiraj_blaaderen, han, leunissen_vutukuri, peng,weiss}   Because of their experimental simplicity, monocomponent colloidal systems have garnered the most attention; however colloidal alloys offer a greater diversity of equilibrium phases and transformation pathways,\cite{casey} though they are more difficult to study.

Here, we study both fluid-solid and solid-solid transformations that occur in a mixture of magnetic and nonmagnetic colloidal particles immersed in a thin aqueous solution of magnetic nanoparticles (ferrofluid) and exposed to an external magnetic field.  This system behaves similarly to a mixture of point dipoles with opposite orientations, where the relative magnitude of the dipole moments can be tuned by changing the ferrofluid concentration.  We study colloidal monolayers by setting the thickness of the fluid layer to be nearly equal to the particle diameter, thus enabling the investigation of transformation pathways with high spatial and temporal precision.  
\begin{figure*}[t]
    \centering
   
   \begin{tabular}{cc}
   	
   	\multicolumn{1}{l}{(a)} & \multicolumn{1}{l}{(d)} \\
   	\includegraphics[width=0.55\textwidth]{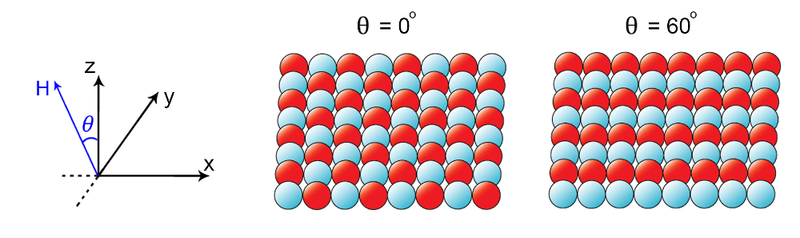} & 
   	\multirow{-6}[2.5]{*}{\includegraphics[width=0.4\textwidth]{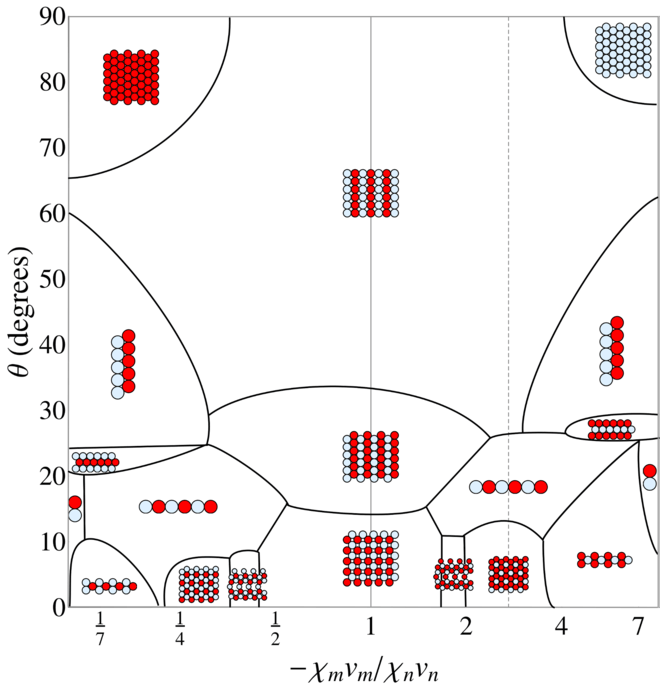}}\\
   	\multicolumn{1}{l}{(b) \hspace{46mm} (c)} \\
    	\includegraphics[width=0.26\textwidth]{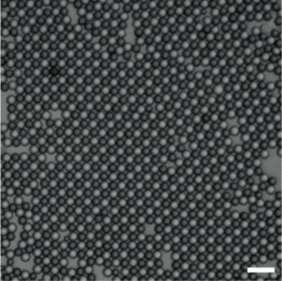}\hspace{5mm}\includegraphics[width=0.258\textwidth]{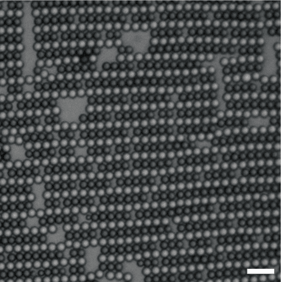}
   \end{tabular}

    \caption{Phases of binary colloidal monolayers. (a) Magnetic particles (red) and nonmagnetic particles (blue) are depicted in the checkerboard crystal phase at low tilt angles and striped crystal phase at high tilt angles.  Experimental images corresponding to the cartoons in (a) are provided in (b) the checkerboard crystal in a tilt angle of $\theta = 0^{\circ}$ and (c) the striped crystal in a tilt angle of $\theta = 60^{\circ}$. Magnetic particles have average diameter of $2.8\,\mu \mathrm{m}$, and nonmagnetic particles have diameter of $3.1\,\mu \mathrm{m}$. Particles are immersed a ferrofluid with 1\% volume fraction of magnetic nanoparticles.  Scale bars are $10\,\mu\mathrm{m}$. (d) Calculated minimal potential energy configuration as a function of tilt angle and negative susceptibility ratio, $-\chi_{\mathrm{m}}v_{\mathrm{m}}/\chi_{\mathrm{n}}v_{\mathrm{n}}$, for a perfectly confined 2D system with equal bead size and particle number densities. Note the natural log scale on the horizontal axis. Gravity and image dipoles were not included in this calculation. The susceptibility ratio can be tuned by changing the nanoparticle concentration in the ferrofluid.  Experiments and simulations in this work were performed at $-\chi_{\mathrm{m}}v_{\mathrm{m}}/\chi_{\mathrm{n}}v_{\mathrm{n}} = 1$ (solid gray line). Note that for $3\lesssim\chi_{\mathrm{m}}v_{\mathrm{m}}/\chi_{\mathrm{n}}v_{\mathrm{n}}\lesssim3.7$ (e.g., dashed gray line), decreasing the tilt angle transforms an equimolar striped crystal into a hexagonal crystal with a 2:1 ratio of magnetic to nonmagnetic particles through an intermediate chain phase. This change in crystal stoichiometry suggests that the solid-solid transition should be accompanied by long-range particle diffusion. }
    \label{fig:Figure1}
\end{figure*}

A rich variety of equilibrium crystal structures can be obtained by adjusting the ferrofluid concentration, the relative particle density $(n_{\mathrm{m}}/n_{\mathrm{n}})$, and the direction of an external field. The minimal energy (zero-temperature) phase diagram computed for 65 different structures (see Fig.~\ref{fig:Figure1}d and Sec. \ref{sec:analytical_phase_diagram} for calculation details) illustrates the wealth of phases formed in an equimolar alloy system ($n_{\mathrm{m}}/n_{\mathrm{n}}=1$). Various solid-solid transformations are achievable, including some diffusionless ones. The experiments reported here roughly follow the solid vertical line of Fig.~\ref{fig:Figure1}d, in which the particle moments are equal and opposite, and where the checkerboard crystal (Fig.~\ref{fig:Figure1}b) transforms to a striped crystal phase (Fig.~\ref{fig:Figure1}c) as the direction of the magnetic field is tilted away from the $z$- direction.

Interestingly, our experimental studies of these transformation pathways indicate the presence of intermediate states that cannot be explained by minimum energy configurations in perfectly confined two-dimensional (2D) monolayers. Our experiments, Monte Carlo simulations, and analytical calculations, further reveal that the specific pathways are highly sensitive to the crystal orientation and to the degree of monolayer buckling allowed, i.e., the distance of separation between the plates that confine the sample. Some of these pathways appear smooth and continuous over a broad range of tilt angles, whereas others proceed via sequences of discrete collective motions.

The rest of this article is organized as follows. Section~\ref{sec:parameters}, describes the important physical features and parameters of the binary colloid system. Sections~ \ref{sec:methods} and \ref{sec:expmethods} detail the numerical and experimental methods, respectively.  Section~\ref{sec:results} presents results from experiments and simulations, including the phase diagram and crystal growth, magnetostrictive response to applied fields for small tilt angles, and the martensitic phase transformation that occurs at large tilt angles.  We briefly conclude with remarks about equilibration times in our experimental system and future research directions.

\section{Theoretical model}
\label{sec:parameters}

The magnetic particles used in experiments consist of magnetic nanograins uniformly distributed inside an inert, spherical micron-sized polymer matrix. They thus behave as a homogeneous continuum on the micron scale, allowing for the assignment of an effective magnetic permeability, $\mu_{\mathrm{m}}$ or $\mu_{\mathrm{n}}$, for magnetic or nonmagnetic particles, respectively.  The nonmagnetic particles have similar composition, but do not contain magnetic nanograins. Both types of particles are immersed in an aqueous ferrofluid of magnetic nanograins, whose concentration tunes the average magnetic permeability $\mu_{\mathrm{f}}$.  In the following, the ferrofluid is assumed to be homogeneous.

\subsection{Dipolar potential energy}
From classical magnetostatics we know that a homogeneous sphere of magnetically susceptible material produces the field of a point dipole when exposed to a uniform magnetic field. The potential energy of the binary colloid system placed in an external magnetic field is thus modeled by taking the particles to be induced point dipoles at the centers of hard spheres, with the effective susceptibility determined by the permeability difference between a sphere and the surrounding fluid. 
The system potential energy $U$ is thus taken to be the sum of hard sphere interactions and magnetic interactions between the induced point dipoles. The gravitational energy associated with out-of-plane buckling is also considered. For $N$ particles, we thus have 
\begin{equation}
	U = \sum_{i<j}^{N} [
U_{\mathrm{HS}}(r_{ij}) +
U_{\mathrm{dd}}(\vec{r}_{ij})+
U_{\mathrm{g}}( \vec{r}_i)]\,
\label{eqn:hamiltonian}
\end{equation}
where $r_{ij}$ is the distance between particles $i$ and $j$, and hard sphere exclusion is complete up to the particle surface at $(\sigma_i+\sigma_j)/2$ for particles of diameter $\sigma$. Dipolar interactions are given by the classical expression,\cite{griffiths}
 \begin{equation}
	U_{\mathrm{dd}}(\vec{r}_{ij}) = -\frac{\mu_{\mathrm{f}}}{4\pi r_{ij}^3}\l[3\,( \vec{m}_i \cdot \hat{\vec{r}}_{ij})( \vec{m}_j \cdot \hat{\vec{r}}_{ij}) - \vec{m}_i \cdot \vec{m}_j\r]\,,
\label{eqn:Udd}
\end{equation}
where $\vec{m}_i$ and $\vec{m}_j$ are the effective dipole moments and ${\vec{\hat r}}$ is a unit vector. The ferrofluid magnetic permeability $\mu_{\mathrm{f}}$ is assumed to be a function of the material bulk magnetic susceptibility $\chi_{\mathrm{B}}$ and the volume fraction of nanoparticles $\varphi$,
\begin{equation}
	\mu_{\mathrm{f}} = \mu_0(1+\varphi\chi_{\mathrm{B}})\,,
\end{equation}
where $\mu_0$ is the vacuum permeability.

The effective magnetic susceptibility of a particle submersed in the ferrofluid is given by
\begin{equation}
	\bar \chi_i = 3 \l( \frac{\mu_i - \mu_{\mathrm{f}}}{ \mu_i + 2 \mu_{\mathrm{f}}} \r)\,,
\end{equation}
leading to the effective dipole moment 
\begin{equation}
	\vec{m}_i = \bar \chi_i  v_i \vec{H}_i\,,
\end{equation}
where $v_i$ is the particle volume. When $\mu_{\mathrm{f}}$ lies between $\mu_{\mathrm{n}}\approx\mu_0$ and $\mu_{\mathrm{m}}>\mu_0$, the nonmagnetic particles are effectively diamagnetic ($\bar \chi_{\mathrm{n}} < 0$), while the magnetic particles are paramagnetic ($\bar \chi_{\mathrm{m}} >0$).

No sufficiently accurate, direct measurements of the difference in susceptibilities between each particle type and the ferrofluid are available. We can, however, estimate the ratio of the two moments in an equimolar mixture by noting that the observed checkerboard phase is a potential energy minimum when the susceptibility ratio $\chi_{\mathrm{m}}v_{\mathrm{m}}/\chi_{\mathrm{n}}v_{\mathrm{n}} = -1$ and the tilt angle $\theta$ is small (see Fig.~\ref{fig:Figure1}d and Sec.~\ref{sec:analytical_phase_diagram}).  In experiments, we focus on systems where the largest checkerboard crystals assemble, and we thus reasonably assume that  $\chi_{\mathrm{m}}v_{\mathrm{m}}/\chi_{\mathrm{n}}v_{\mathrm{n}} = -1$ for the rest of the experimental and numerical analyses.

\subsection{Self-consistent magnetic moments and image dipoles}
\label{sec:moments}
The magnetic moment of a given particle should be calculated from the total field at its center, which is the sum of the applied field and the field created by neighboring particles:
\begin{equation}
\vec{m}_i = \bar \chi_i v_i \l [ \frac{\mu_0}{\mu_{\mathrm{f}}} H_z{\bf \hat z} + H_x {\bf \hat x}  + \sum_{j \in\partial i^{(\xi)} } \frac{3(\vec{m}_j \cdot{\vec{\hat r}}_{ij}){\vec{\hat r}}_{ij} - \vec{m}_j}{4\pi r_{ij}^3} \r]\,,
\label{eqn:moments}
\end{equation}
where $H_z$ ($H_x$) is the vertical (in-plane) component of the external field in air and $\partial i^{(\xi)}$ denotes the set of neighbors of particle $i$ within a cutoff radius $\xi$.

\begin{figure*}
	\centering
	\includegraphics[width = 1\textwidth]{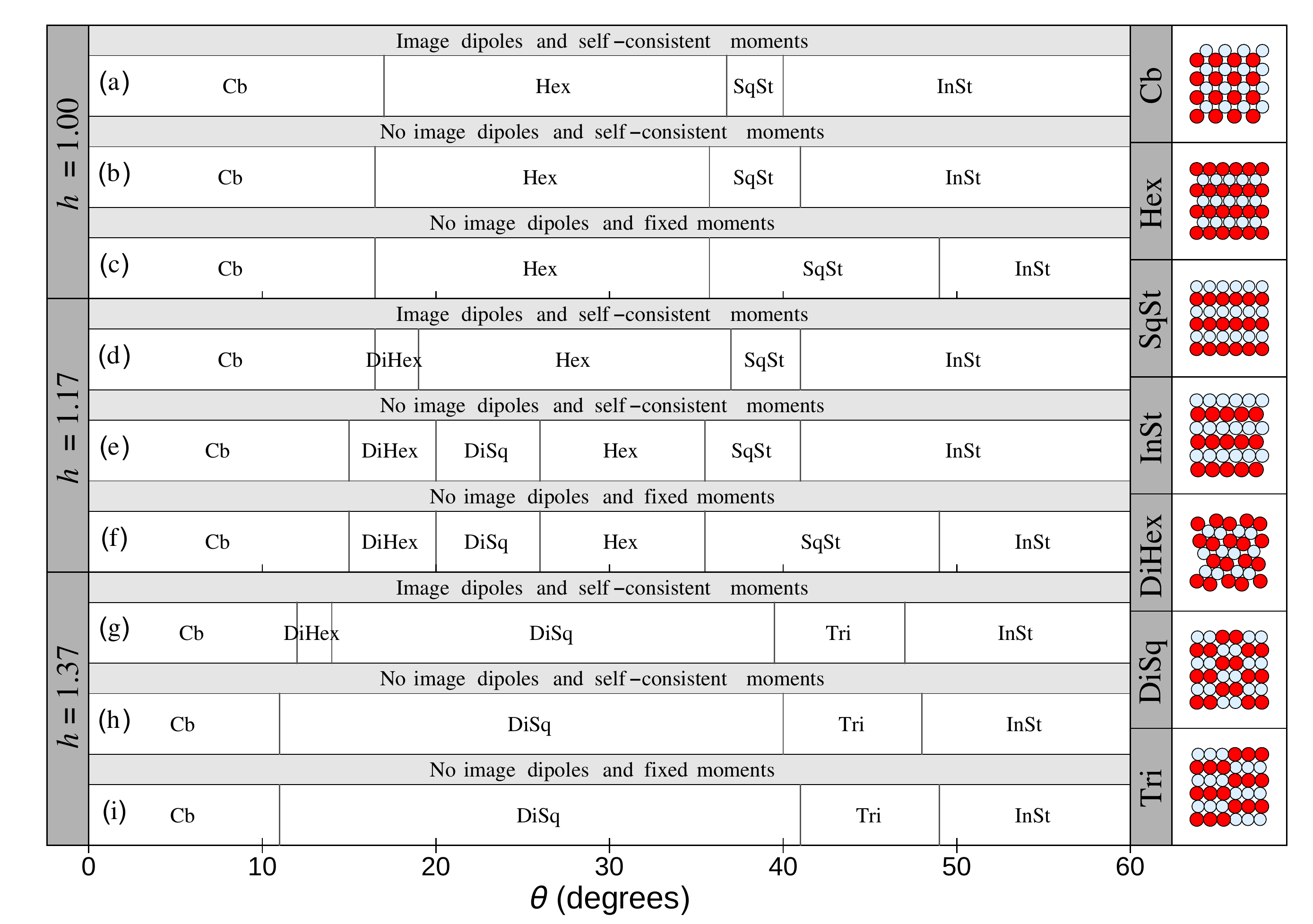}
	\caption{ Minimal energy structures at various field tilt angles $\theta$. Energies were calculated using different values of glass slide to coverslip distance $h$, and $h=1.00$ corresponds to a perfect 2D confinement. For each buckling height, the effects of image dipoles and self-consistently determined magnetic moments were included in the energy calculation. In panels (a), (d), and (g) image dipoles were included and self-consistently determined moments were used. In panels (b), (e), and (h) no image dipoles were included and self-consistently determined moments were used. In panels (c), (f), and (i) no image dipoles were included and the magnetic moments were fixed by the external field alone. }
\label{fig:Figure2}
\end{figure*}

Such computations are possible, but numerically expensive, so we first evaluate their relevance. Self-consistent calculations of the effect of nearest-neighbor fields on the magnetic moments reveal that they generate only small contributions to the potential energy. The only structure with a non-negligible change in potential energy is the incommensurate stripe phase, but even in this case the zero-temperature phase boundary shifts by only a few degrees, from $41^{\circ}$ to $49^{\circ}$ (see Fig.~\ref{fig:Figure2}). For computational efficiency, we therefore ignore the field created by other particles when calculating the magnetic moments in Monte Carlo simulations.

The presence of a magnetic permeability mismatch at the fluid-glass interface gives rise to image dipoles.  The difference in magnetic permeabilities of the ferrofluid and the confining glass results in an additional field felt by the particles.  Consider a point  dipole with magnetic moment $\vec m = (m_x, m_y, m_z)$ located at $\vec r = (x_0, y_0, z_0)$ within the ferrofluid. Let the bottom glass slide be in the plane $z = 0$ and the coverslip be in the plane $ z = h$.  The field within the ferrofluid at $\vec r = (x, y, z)$ is thus a sum of the magnetic field of the real dipole and the fields of two image dipoles located at $\vec r_{\mathrm{im}}^{(1)} = (x_0, y_0,  2h-z_0)$ and $\vec r_{\mathrm{im}}^{(2)} = (x_0, y_0,  -z_0)$ with magnetic moment:\cite{jackson}
\begin{equation}
\vec m_{\mathrm{im}} = \l( \frac{\mu_f - \mu_0}{\mu_f + \mu_0} \r) (m_x, m_y, -m_z)\,.
\end{equation}

The inclusion of image dipoles in potential energy calculations produces slightly different minimum energy phase boundaries, but again the phase transition sequence is not qualitatively affected (Fig.~\ref{fig:Figure2}), which justifies ignoring their effect in simulations.

\subsection{Gravitational effects}
A gravitational contribution must be included when the experimental cell height is larger than one particle diameter.  For a density mismatch $\bar \rho_{\mathrm{i}}$ between particles of type $i$ and the ferrofluid, we take
\begin{equation}
U_{\mathrm{g}}(\vec{r}_i) = \bar\rho_{\mathrm{i}} v_{\mathrm{i}} g\, \vec{r}_i \cdot \vec{ \hat z}\,,
\label{eqn:Ug}
\end{equation}
where $v_{\mathrm{i}}$ is the volume of a bead of type $i$ and the $z = 0$ plane is at a distance $\sigma_{\mathrm{m}}/2$ above the bottom glass slide.  In our experiments, the density of the nonmagnetic beads is closely matched with that of the fluid, but the magnetic beads are more dense, hence: $\bar \rho_{\mathrm{n}} = 0$ and $\bar \rho_{\mathrm{m}} = 350\,\mathrm{kg}/\mathrm{m}^3$.

\subsection{Reduced temperature}
In order to directly compare simulations, in which the temperature is varied, with experiments, in which the applied field is controlled, we define a reduced temperature as the ratio of thermal energy to dipolar potential energy in the system:
\begin{equation}
T^* \equiv \frac{4\pi k_{\mathrm{B}} T\alpha^2}{\mu_0 H^2 \sigma^3}\,,
\label{eqn:reducedT}
\end{equation}
where $T=298\,\mathrm{K}$ is the ambient temperature, $k_{\mathrm{B}}$ is the Boltzmann constant, $\sigma=(\sigma_{\mathrm{m}}+\sigma_{\mathrm{n}})/2$ is the average particle diameter, and $\alpha$ is an effective susceptibility that accounts for experimental features that are not modeled directly and corrections to the dipole moment approximations mentioned above.

\section{Numerical Methods}
\label{sec:methods}
Different methods were used to extract quantitative and qualitative information from the variants of the theoretical model. The numerical and simulation details are provided in this section.

\subsection{$T=0$ phase diagram}
\label{sec:analytical_phase_diagram}

The minimal potential energy structures as a function of $\theta$ and $-\chi_{\mathrm{m}}v_{\mathrm{m}}/\chi_{\mathrm{n}}v_{\mathrm{n}}$ were determined from a set 65 two-dimensional structures of equal-sized magnetic and nonmagnetic particles, i.e., $\sigma_{\mathrm{m}}=\sigma_{\mathrm{n}}$. This set includes rings, chains and crystals that were either presented in Ref.~\onlinecite{khalil}, manually constructed, or predicted by a genetic algorithm.\cite{filion} Potential energy and magnetic moment calculations were performed using the same methods as in Ref.~\onlinecite{khalil}.  Magnetic moments were self-consistently determined by solving Eq.~\eqref{eqn:moments} using a cutoff radius of $\xi = 1.1\,\sigma$. The overall potential energy calculation used no cutoff radius for the rings and cutoff radii of $500\,\sigma$ and $100\,\sigma$ for the chains and crystals, respectively.  Gravity and image dipoles were neglected, and, for non-stoichiometric crystals, additional particles were assigned zero potential energy, as in the ideal gas limit.\cite{khalil}  Other parameters used in the calculations are presented in Table~\ref{tab:theory_parameters}. Note that performing the energy calculations with different values of $\sigma$, $|\vec H|$, and $\chi_{\mathrm{B}}$ does not change the location of the phase boundaries in the tilt angle-susceptibility ratio phase diagram.

\begin{table}[ht]
\begin{center}
\begin{tabular}{  | c | c | c | }
\hline
 & {\bf Fig.~\ref{fig:Figure2}} &  {\bf Fig.~\ref{fig:Figure1}d }\\ \hline
\hline
	$\sigma_{\mathrm{m}}$ & $2.8\, \mu\mathrm{m}$ &  $1.0\, \mu\mathrm{m}$\\ \hline
	$\sigma_{\mathrm{n}}$ & $3.1\, \mu\mathrm{m}$ &  $1.0\, \mu\mathrm{m}$\\ \hline
  $\mu_{\mathrm{m}}$ & 1.5 & 2\\ \hline
	$\mu_{\mathrm{n}}$ & 1 & 1\\ \hline
$\chi_{\mathrm{B}}$ & $19.5$ & $11.65$\\ \hline
$|\vec{H}_0|$ & $ 12\, \mathrm{Oe}$  & $ 50\, \mathrm{Oe}$\\ \hline
$\varphi$ & $ 1.0\, \%$ & -\\ \hline
$\psi$ & $ 0^{\circ}$ & -\\ \hline
$\xi $ & $ 20.1 \, \sigma_{\mathrm{n}}$ & $1.1\,\sigma$ \\ \hline
$r_{\mathrm{u}}$ & $ 30.1\, \sigma_{\mathrm{n}} $ &see Sec.\ref{sec:analytical_phase_diagram}\\
 \hline
	\end{tabular}
\end{center}
\caption{Parameters used in potential energy calculations for Fig.~\ref{fig:Figure1}d and Fig.~\ref{fig:Figure2}.}
\label{tab:theory_parameters}
\end{table}

Comparisons of potential energies with and without self-consistent fields and image dipoles were made for seven types of structures. Here again, energy calculations were performed in the high-field limit, in which the gravitational energy is negligible. Three buckling heights were then considered, corresponding to slide to coverslip separations  $h = \sigma_{\mathrm{n}}$,  1.17  $\sigma_{\mathrm{n}}$, and 1.37  $\sigma_{\mathrm{n}}$.

For calculations in which magnetic moments were self-consistently determined, Eq.~\eqref{eqn:moments} was solved using the Jacobi method with a cutoff radius of $\xi = 20.1 \,\sigma_{\mathrm{n}}$. Convergence of the energy to within 0.1\% was obtained within three iterations. For all cases, the total potential energy per particle was calculated using a cutoff radius of $r_{\mathrm{u}} = 30.1\, \sigma_{\mathrm{n}}$.  The two cutoff radii were chosen so as to minimize computation time while yielding an error of $\le 0.1\%$ in the potential energy. The values of the parameters used in calculations are presented in Table~\ref{tab:theory_parameters}.

The values of $\varphi$ and $\chi_{\mathrm{B}}$ were chosen such that the magnetic moments of the magnetic and nonmagnetic particles are equal and opposite in the limit of infinite dilution.  This results in the checkerboard crystal being stable at $\theta = 0$ (Fig. 1d). Note that because both the magnetic moments and the dipole energy are functions of the product $\varphi \chi_{\mathrm{B}}$, calculations performed with any values of $\varphi$ and $\chi_{\mathrm{B}}$ that satisfy the magnetic moment condition above result in identical system energies.

The value of $\mu_{\mathrm{m}}$ in experiment is not known precisely.  For the structures observed experimentally (depicted in Fig.~\ref{fig:Figure2}), energy calculations were performed with $\mu_{\mathrm{m}} = 1.5$, which accounts well for the tilt angles at which transitions should occur.

\subsection{Finite $T$ simulations}

The phase diagram and hysteresis loop calculations were obtained from systems in perfect 2D confinement. For simplicity and efficiency, two additional approximations were made. First, the magnetic and nonmagnetic particles were chosen to have the same diameter, $\sigma$. Second, the magnetic moments were not calculated self-consistently, but kept fixed (see Sec.~\ref{sec:moments} for a detailed discussion). The first approximation was only made for the phase diagram calculation and the second one was made for all simulations. It was empirically observed from comparing the results with experiments that the errors introduced by these approximations can be accounted for by rescaling the fitting factor $\alpha$.

In order to keep the error on the numerical energy calculations within 0.1\%, the cutoff radius for the pair interaction $r_{\mathrm{sim}}$ was chosen to be half the simulation box size, i.e.,  a given particle is considered to interact with $\sim N\pi/4$ particles. To further improve the computational efficiency, energy calculations for the solid phase used Ewald summation (see, e.g. Ref.~\onlinecite{lipkowitz}, Ch.~6,~App.~F).  For the fluid phase, however, in the system size regime considered the approach is less efficient than real-space radial truncation.

\subsection{Equations of state}

The fluid and the crystal equations of state (EOS) were determined by measuring density $\varrho$ as a function of pressure $P$ from constant $NPT$ Monte Carlo (MC) simulations. Simulations with  $N=400$ particles were averaged over $4\times10^6$ MC cycles following an initial equilibration of $10^6$ MC cycles. Each MC cycle consists of $N$ particle moves and one volume move, on average. In order to efficiently treat strongly associated particles at low densities ($\varrho \lesssim 10^{-2}$) and temperatures ($T\lesssim0.14$), aggregation-volume-bias MC (AVBMC)\cite{chen_B} and cluster volume moves\cite{almarza} are substituted for half of the standard local displacements and volume moves. We used $r_{\mathrm{AVBMC}} = 1.2\sigma_{ij}$ as the cutoff for the AVBMC and $r_{\mathrm{clu}} = 1.05\sigma_{ij}$ as the cutoff for the cluster volume move, where $\sigma_{ij}=(\sigma_{\mathrm{i}}+\sigma_j)/2$. The maximal step size $\Delta x$ and $\Delta \ln V$ for displacements and volume moves, respectively, were tuned every 1000 MC cycles, in order to keep the acceptance rates of both types of moves between 30\% and 40\%.

\subsection{Phase diagram}

The system free energy was obtained through thermodynamic integration (see, e.g. Ref.~\onlinecite{frenkel}, Ch.~10). For the fluid, an equimolar mixture of ideal gases is used as the reference free energy 
\begin{equation}
F^{\mathrm{id}}(\varrho)=Nk_{\mathrm{B}}T\left[\ln\left(\frac{\varrho \Lambda^{2}}{2}\right) - 1 + \frac{\ln(2\pi N)}{2N}\right],
\end{equation}
where the thermal de Broglie wavelength $\Lambda$ is here set equal to $\sigma$. For the solid, an Einstein crystal with area fraction $\eta=0.739$ was used instead. We found that the Einstein crystal limit is recovered for a spring constant of $2000 k_B T/\sigma^2$ for the different state points studied.

The fluid-solid coexistence densities at a given $T$ were determined from a parametric curve of the chemical potential, $F/N+P/\varrho$, and $P$ for each phase as a function of density. Phase coexistence takes place at the intersection of the two curves. Note that the fluid-checkerboard crystal coexistence region obtained here is consistent with an earlier, lower-precision calculation.\cite{Ito}

\subsection{Dynamical Monte Carlo} 

The simulation results for both the hysteresis loop and the martensitic transformation, dynamical Monte Carlo simulations, i.e., using only local single-particle displacements, were performed at constant $NVT$ and for $N=400$.

The hysteresis loop of Fig.~\ref{fig:Figure4} was determined at a fixed area fraction $\eta=0.46$ in a 2D system. The order parameter was averaged over $2000$ MC cycles for each $T$ ($H$). The reduced temperature was changed by $\Delta T=0.004$ every 2000 MC cycles between $T^{*}=0.023-0.075$, $\Delta T=0.015$ every 2000 MC cycles between $T^{*}=0.075-0.135$, and $\Delta T=0.03$ every 2000 MC cycles between $T^{*}=0.135-0.375$.

For the martensitic transformation simulation, the full Hamiltonian for particles with diameter ratio $\sigma_{\mathrm{m}}:\sigma_{\mathrm{n}}=2.8:3.1$ (Section~\ref{sec:parameters}) was used.  Both perfectly 2D and quasi-2D systems were considered. In the former, particle centers were constrained to only move in the $z=\sigma_{\mathrm{n}}/2$ plane; in the latter, the particles were allowed to move between hard walls separated by a distance  $h = 1.11\sigma_{\mathrm{n}}$.  The external field was tilted from $0^{\circ}$ to $50^{\circ}$ with a tilt rate of $2.5^{\circ}/2000\ \mathrm{MC\ cycles}$. To straightforwardly compare the simulation results with the experiments, images and movies were colored using the same protocol as used in experiments (See Section~\ref{sec:phi_4} and Movie S4).

\section{Experimental Methods}
\label{sec:expmethods}
Colloidal monolayers were formed by placing an equimolar (1:1) binary mixture of spherical $2.8\,\mu\mathrm{m}$ diameter magnetic (M-270 Dynabeads$^{\circledR}$, Life Technologies$^{\mathrm{TM}}$) and $3.1\,\mu\mathrm{m}$ nonmagnetic particles (Fluro-Max R0300, Thermo Fischer) immersed in a ferrofluid (EMG705, Ferrotec, Bedford, NH) between a glass slide and a coverslip.  Both magnetic and nonmagnetic particles have a size dispersity of less than 3\%.   To reduce adhesion between particles and surfaces, the glass slides and coverslips were coated with polyethyleneglycol (PEG) by silanization with 10kD silane-polyoxyethylene-carboxylic acid (PG2-CASL-10k, NANOCS, New York, NY).  The ferrofluid susceptibility was controlled by adjusting the volume fraction of magnetic nanoparticles in the final suspension to $\sim 1\%$.

A $1.9 \, \mu\mathrm{L}$ aliquot of fluid mixture was sealed in between the glass surfaces with Loctite marine epoxy. Uniform magnetic fields were applied by passing currents through air-core solenoids (Fisher Scientific, Pittsburgh, PA).  Microscopy was performed with a DM LM fluorescent microscope (LEICA, Bannockburn, IL) using a 40X air-immersion objective and a combination of brightfield and fluorescent filter cubes (Chroma Technology).  Videos were recorded with Retiga 2000R camera (Qimaging, Surrey, Canada).  The field ramp was produced by a combination of vertical and horizontal solenoids controlled with Labview (National Instruments, Version 2010, Austin, TX).   Image processing was performed using MATLAB (Mathworks, Version 2011, Natick, MA).  Details about the image processing and data analysis methods\cite{chen_B, almarza} are provided in Section~\ref{sec:phi_4}.

\subsection{Phase diagram}

The fluid-solid transition was studied in a vertical magnetic field, with samples prepared at various $\eta$. (See Fig.~\ref{fig:Figure5}.)  
Each cooling-heating cycle lasted 512 minutes. The magnetic field was adjusted with the square root of time, which results in the inverse reduced temperature changing linearly with time.  The time-dependence of the magnetic field over a single cycle was thus
\begin{equation}
	\vec{H} = \begin{cases}  0.5 \sqrt{t}\,\,{\vec{ \hat z}}, &0 \le t \le 256 \\ 0.5\sqrt{512-t}\,\,{ \vec{\hat z}}, &  256 \ge t \le 512\,,\end{cases}
\end{equation}
where the field strength is in Oe and time is in minutes. An image of the system was taken every minute. An example of the fluid-solid transition in an effective cooling-heating cycle at $\eta=0.43$ is demonstrated in Movie S1.

\subsection{Magnetostriction}

Magnetostriction experiments proceed by adding a weak, in-plane magnetic field to a vertical field of $9.5\,$Oe, hence the total external magnetic field was 
\begin{equation}
	\vec{H}_0 =  H_z\,{\vec{\hat z}} + H_x \sin( 2 \pi f t)\, {\vec{\hat x}}\,,
\label{eqn:fixedField}
\end{equation}
where $H_z = 9.5\,$Oe,  $H_x = 2.1\,$Oe. The cycling frequency  $f = 5.0\times10^{-4}$ Hz was chosen such that the crystal had sufficient time to equilibrate, and it corresponds to the polar field tilt angle reaching $12.5^{\circ}$ from $0^{\circ}$ in $2.00\times10^3$ s.

The magnetostriction dependence on the polar tilt angle  (Fig.~\ref{fig:Figure6}d) was determined from a square section of the system consisting of $6\times6$ nonmagnetic particles within a crystallite (Fig.~\ref{fig:Figure6}b). In order to reduce the influence of other factors, such as crystal size and orientation, we only analyzed crystals containing more than 250 particles and with their 10 axis parallel to the $x$-axis (azimuthal angle $\psi = 0^{\circ}$). During each of the eight replicates, an image of the system was taken every 50 s, and their analysis gave consistent results.

The magnetostriction dependence on the crystal orientation, or azimuthal angle $\psi$, was determined from crystallites with more than 250 particles. Crystals with similar orientation ($\Delta\psi<5^{\circ}$) were grouped together and crystal growth was repeated until every group had at least five data sets. Within each group, the average relative extension ratio at $\theta = 12.5^{\circ}$ was calculated (Fig.~\ref{fig:Figure6}e) from images of the system taken every 20 s. An example of magnetostriction dependence on both $\theta$ and $\psi$ is presented in Movie S3.

\subsection{Martensitic transformation}

More than 10 martensitic transformation experiments were performed to verify the consistency of the different martensitic transformation pathways, and an image was taken every 30 s for analysis in each experiment. The image processing and particle identification protocol described in Section~\ref{sec:phi_4} was used to determine the particle coordinates. The particle bonds were then characterized using the method described in Section \ref{sec:particle_bonds}. In Figure~\ref{fig:Figure7} and Movies S8, S10, S12, and S13, the particles are artificially colored according to their bond type for visualizing the details in the martensitic transformation process.

\subsection{Modified square order parameter }
\label{sec:phi_4}

Experimental measurements of $\Phi_4$ were obtained from images taken during the cooling-heating cycle that were preprocessed using background subtraction and noise reduction, following the method developed by Crocker et al.\cite{crocker}  A two-step particle identification was then performed: (i) nonmagnetic particles were identified by finding the local intensity maximum of the particle center; (ii) magnetic particles were identified by first locating the dark rim of the particle edge, and then checking that the intensity of the tentative particle center is within a properly chosen range. Visual inspection of multiple images reveals that this identification protocol correctly determines $>98\%$ of the particle coordinates.  For Movies S8, S10, S12, and S13, the remaining particles were manually corrected.

\begin{figure}[t]
  
  \centering
   \begin{tabular}{c}
   	
   	\multicolumn{1}{l}{(a) \hspace{23mm} (b) \hspace{23mm} (c)}  \\
   	\includegraphics[width=0.32\columnwidth]{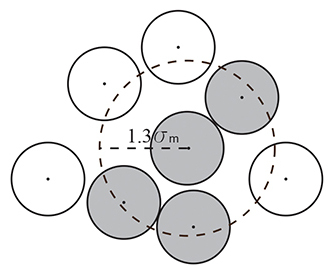}  
    	\includegraphics[width=0.32\columnwidth]{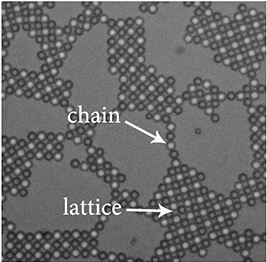}  
    	\includegraphics[width=0.317\columnwidth]{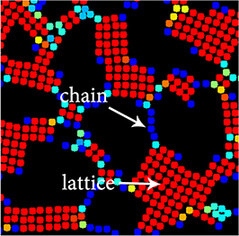} \\

   \end{tabular}
	\caption{ Illustration of the order parameter calculation. (a), Particles within $1.3\sigma_{\mathrm{m}}$ from the particle of interest are considered to be neighbors and are colored gray; particles beyond that distance are colored white.  An example experimental configuration (b), gives a corresponding order parameter analysis (c).  Particles are colored according to their individual values of $\Phi_4$ (Section 4.4), using the color scheme shown in Figure 4b.}
	\label{fig:Figure3}
\end{figure}

Neighbors, i.e., particles within $1.3  \sigma_{\mathrm{m}}$ of each other, are then identified. This definition deliberately excludes second-nearest neighbors in a square lattice. Given a particle $k$ with $n$ neighbors, the local order around particle $k$ is analyzed in terms of a modified square order parameter
\begin{equation}
	\Phi_{4,k} \equiv \begin{cases} 
 0, & n\le 2 \\ 
\frac{1}{n}\l | \sum \limits_{j\in \partial k^{(\Phi)}} \exp(4 i \theta_{j,k})\r |, & n>2\,, 
\end{cases}
	\label{eqn:orderParameter}
\end{equation}
where $\partial k^{(\Phi)}$ denotes the set neighbors (within $r_{\Phi} = 1.3\sigma_{\mathrm{m}}$ of particle $k$), and $\theta_{j,k}$ is the acute angle between the $x$ axis and the bond between particle $k$ and its $j^{\mathrm{th}}$ neighbor. This definition allows us to differentiate between the checkerboard crystal phase ($n=4$ in the bulk, $n=3$ at the boundary, hence $\Phi_{4,k}\neq0$)  and the chain phase ($n=2$ in the bulk, $n=1$ at the boundary, hence $\Phi_{4,k}=0$). An illustration of this calculation is given in Fig.~\ref{fig:Figure3}.
The global order parameter is then defined as the average of $\Phi_{4,k}$ over all particles,
\begin{equation}
	\Phi_4 \equiv \frac{1}{N} \sum_{j=1}^N \Phi_{4,k}\,,
\end{equation}
where $N$ is the total number of particles in the field of view.

\subsection{Particle bond characterization}
\label{sec:particle_bonds}

Types of particle bonds, rather than a local order parameter, were used to quantify the martensitic transformation. When $r_{ij}<1.2\sigma_{ij}$, two particles are deemed bonded. If the bond formed by a pair of like particles lies within $20^{\circ}$ from the in-plane field component, then it is considered to be a martensitic bond; if a bond is between a pair of unlike particles, then it is considered to be an austenitic bond. The number and distribution of bonds is used to characterize the formation of phases in the martensitic transformation process. 

\subsection{Reciprocal lattices }
\label{sec:reciprocal lattices}

The reciprocal lattices were obtained by computing the Fourier transform of experimental and simulation images. To eliminate artifacts created by the image boundary, the real-space images were first processed with a Gaussian kernel function $\exp\{-[(x-x_0)^2+(y-y_0)^2]/2\sigma^2\}$, where $(x_0, y_0)$ is the center of the image and $\sigma = 1000$. A Fast Fourier Transform (FFT) was then performed on the processed images using Matlab (R2011b).

\section{Results}
\label{sec:results}

Results for assembly and phase transformation of the experimental and the theoretical models are presented in this section.

\subsection{Phase diagram in a vertical field}
\label{sec:phase_diagram}

To study the transformation between bulk solid phases without interference from edge effects and defect-mediated transitions, we first developed a protocol for building large, well ordered single crystals.  This endeavor requires a quantitative understanding of the fluid-solid phase transition, which is achieved by directly comparing experimental results with the phase diagrams obtained by Monte Carlo simulations.

\begin{figure}[t!]

   \centering
   \begin{tabular}{l}
   	
   	\multicolumn{1}{l}{(a)}\\
   	\includegraphics[width=0.78\columnwidth]{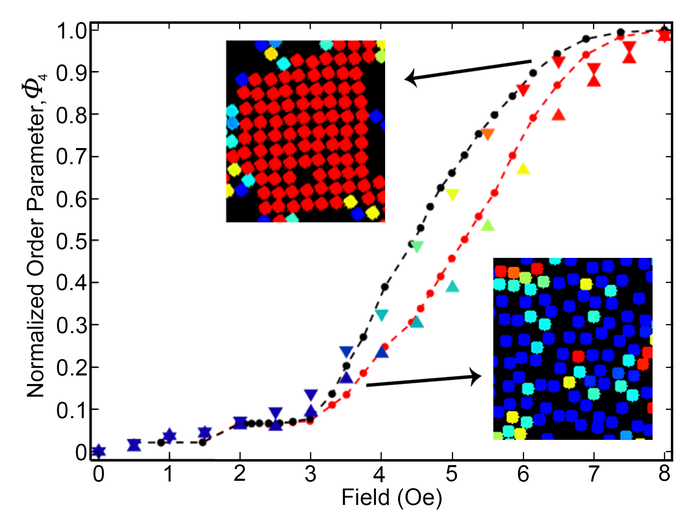} \\
   	\multicolumn{1}{l}{(b)}\\
   	\includegraphics[width=0.95\columnwidth]{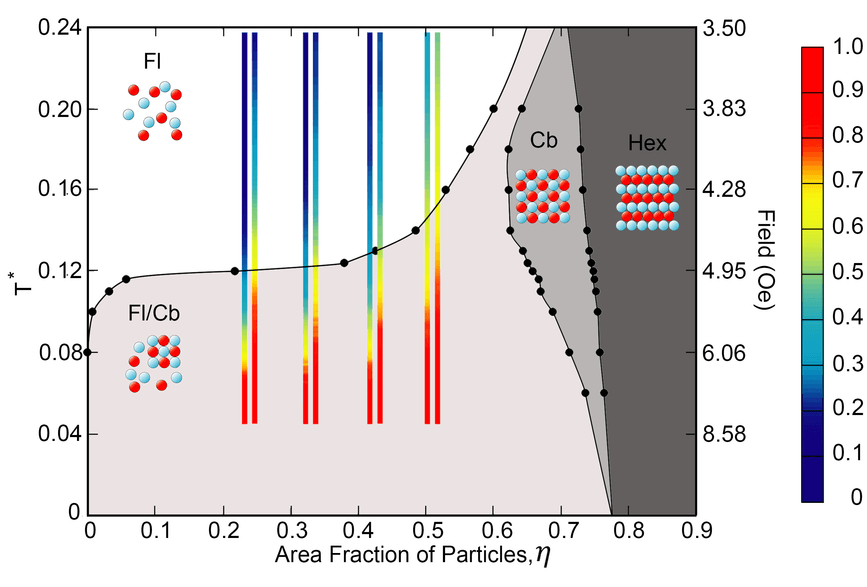}
   	
   \end{tabular}

    \caption{ Fluid-solid phase transitions of binary colloidal monolayers in a vertical field. (a) Experimental and simulated hysteresis curves. Up (down) pointing triangles correspond to experimental data, while the red (black) dashed curve corresponds to the numerical simulation results for the increasing (decreasing) magnetic field. The triangle colors represent the $\Phi_4$ bond orientation order parameters, following the color scheme used in (b). Simulation results (dots connected with dashed lines) are superimposed. (insets) Typical experimental images with each particle colored according to its individual $\Phi_4$ value. Hysteresis curves are obtained with a field ramp rate that grows with the square-root of time, corresponding to a linear temperature ramp. (b) Phase diagram from simulations shows the fluid (Fl), fluid-checkerboard crystal coexistence (Fl/Cb), checkerboard crystal (Cb), and hexagonal striped crystal (Hex) phases. Black dots and connecting lines indicate the simulated phase boundaries.  The colored bars represent the system $\Phi_4$ bond order parameter taken from the heating/cooling cycles of part (a) at four different area fractions. The left and right colored bars at each particle concentration represent the cooling and heating processes, respectively. Particles are immersed in a ferrofluid with 1\% volume fraction of magnetic nanoparticles. The numerical fitting parameter is $\alpha = 2.4\pm0.3$. The color scheme is the same as Fig.~\ref{fig:Figure3}.}
    \label{fig:Figure4}
\end{figure}

Figure~\ref{fig:Figure4}b shows the simulated phase diagram in the plane of reduced temperature $T^*$ and area fraction $\eta$ for dipolar hard spheres under 2D confinement.  The corresponding experimental results are obtained by continuously varying the magnitude $H$ of a vertical magnetic field for samples with different $\eta$. A bond-orientational order parameter $\Phi_4$ (defined in Section~\ref{sec:phi_4}) tracks the first-order freezing and melting transitions as the field strength is cycled between 0 and 8 Oe (Fig.~\ref{fig:Figure4}a and Movies S1 and S2). The center of the hysteresis loop is used to estimate the melting point at a given $\eta$. In agreement with simulations, the melting field strength is roughly independent of $\eta$ at $0.1<\eta < 0.4$, as evidenced by the plateau of Fig.~\ref{fig:Figure4}b, but decreases at high $\eta$. Matching the experimentally observed transitions to the computed phase diagram at several different area fractions gives a fitting factor $\alpha = 2.4 \pm 0.3$, for Eq.~\eqref{eqn:reducedT}

\begin{figure}[t!]
    \centering
    
    \begin{tabular}{l}
	\multicolumn{1}{l}{\hspace{1mm} (a) \hspace{23mm} (b) \hspace{20mm} (c)}  \\
	\includegraphics[width=0.34\columnwidth]{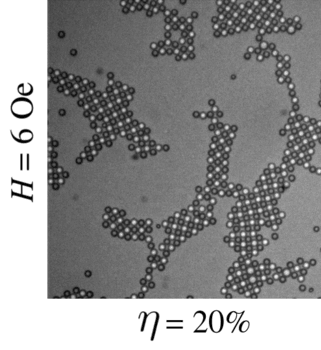}  
	\includegraphics[width=0.29\columnwidth]{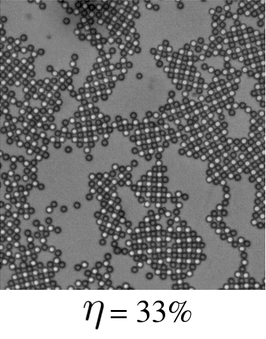}  
	\includegraphics[width=0.293\columnwidth]{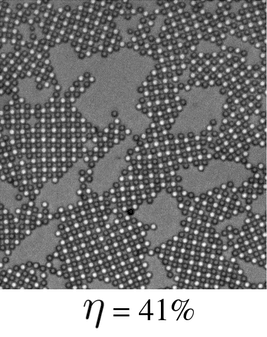} \\
	\multicolumn{1}{l}{\hspace{1mm} (d) \hspace{23mm} (e) \hspace{20mm} (f)}  \\
	\includegraphics[width=0.344\columnwidth]{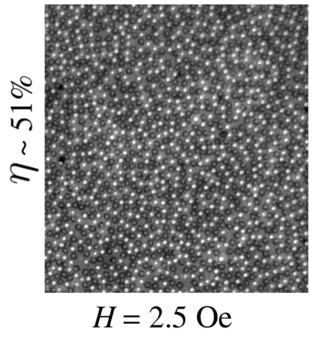}  
	\includegraphics[width=0.29\columnwidth]{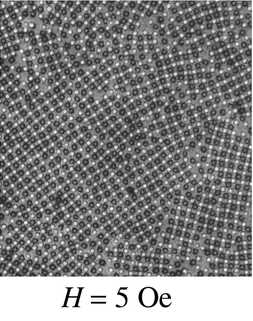}  
	\includegraphics[width=0.293\columnwidth]{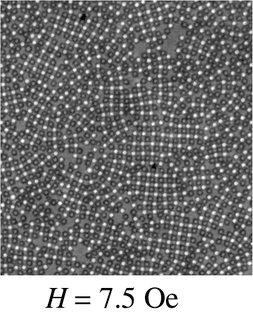} \\
	
    \end{tabular}

    \caption{Crystal growth in a constant vertical field. The dependence of crystal growth on area fraction is demonstrated for (a) $\eta=0.20$, (b) $\eta=0.33$, and (c) $\eta=0.41$ at a field strength of $H=6$ Oe. The dependence of crystal growth on field strength is demonstrated for (d) $H=2.5$ Oe, (e) $H=5$ Oe, and (f) $H=7.5$ Oe at a particle area fraction of $\eta \sim 0.51$. In each experiment, the magnetic field was held constant for six hours. The largest crystals were found to grow at $\eta \sim 0.51$ and $H = 5\,$Oe. 
}
    \label{fig:Figure5}
\end{figure}

In the fluid-solid coexistence regime, we expect that large regular crystals should form at values of $H$ and $\eta$ near the neck of the phase diagram around $T^* = 0.12$. The small density difference between the two phases in that regime helps anneal crystal defects. Preparing monolayers with $\eta \ge 0.60$ was too difficult to achieve because of the high viscosity of these solutions, but reasonably large crystals were nonetheless grown near the phase boundary at $\eta = 0.51$ and $H = 5$ Oe (Fig.~\ref{fig:Figure5}e). Crystals with more than 500 particles, such as those in Fig.~\ref{fig:Figure1},  formed within hours and were taken as the starting points for studies of magnetostriction and martensitic transformations.

\subsection{Tilting the field at small angles: magnetostriction}
\label{sec:magnetostriction}

\begin{figure*}[t]
    \centering
    
    \begin{tabular}{c}
	\multicolumn{1}{l}{\hspace{10mm} (a) \hspace{40mm} (b) \hspace{42mm} (c)}  \\
	\hspace{5mm}\includegraphics[width=0.2\textwidth]{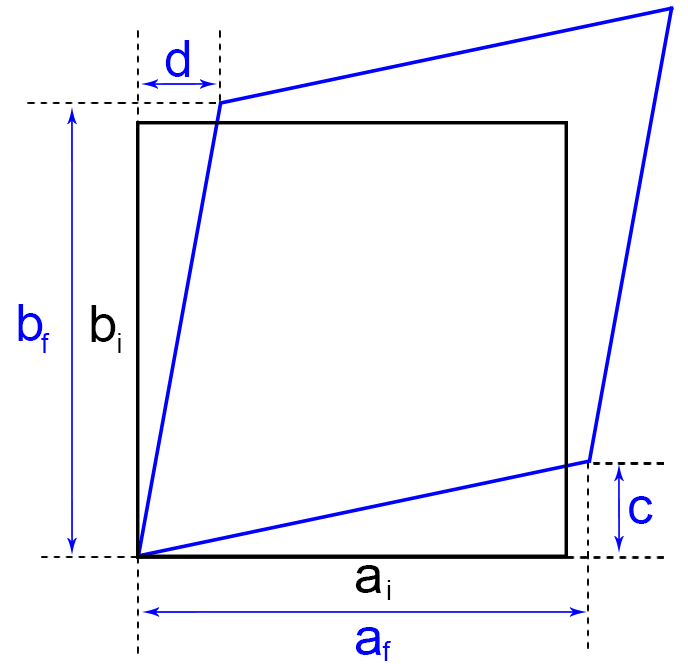}  \hspace{10mm}
	\includegraphics[width=0.2\textwidth]{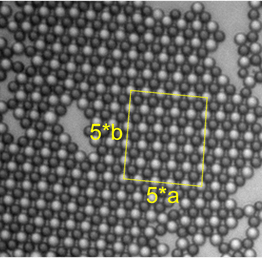} \hspace{10mm}
	\includegraphics[width=0.195\textwidth]{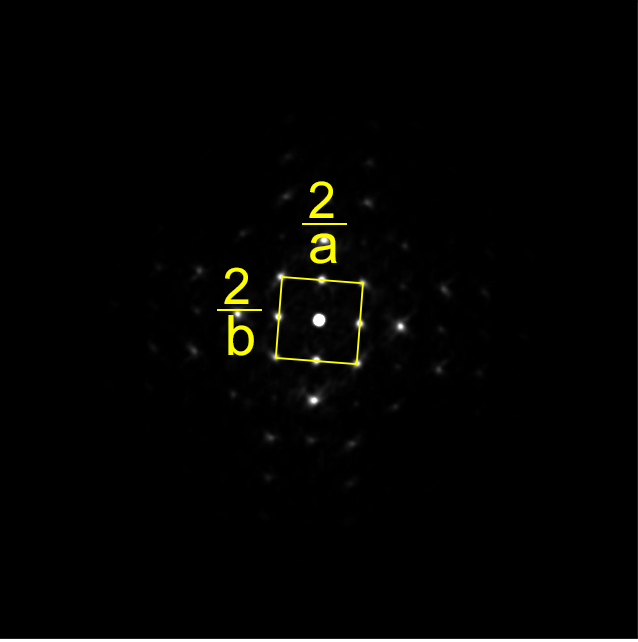} \\
	\multicolumn{1}{l}{\hspace{10mm} (d) \hspace{40mm} (e) \hspace{42mm} (f)}  \\
	\includegraphics[width=0.26\textwidth]{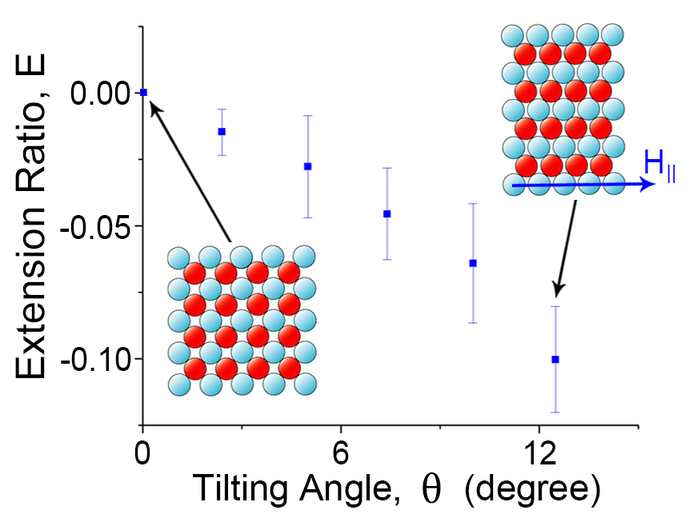}  
	\includegraphics[width=0.26\textwidth]{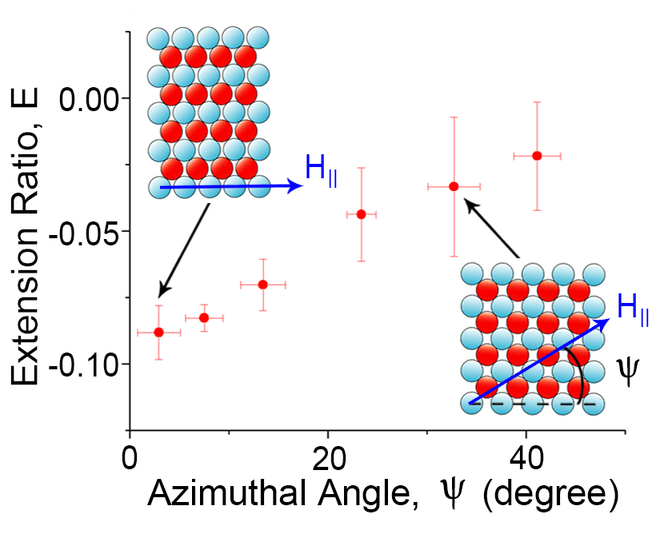}  
	\includegraphics[width=0.26\textwidth]{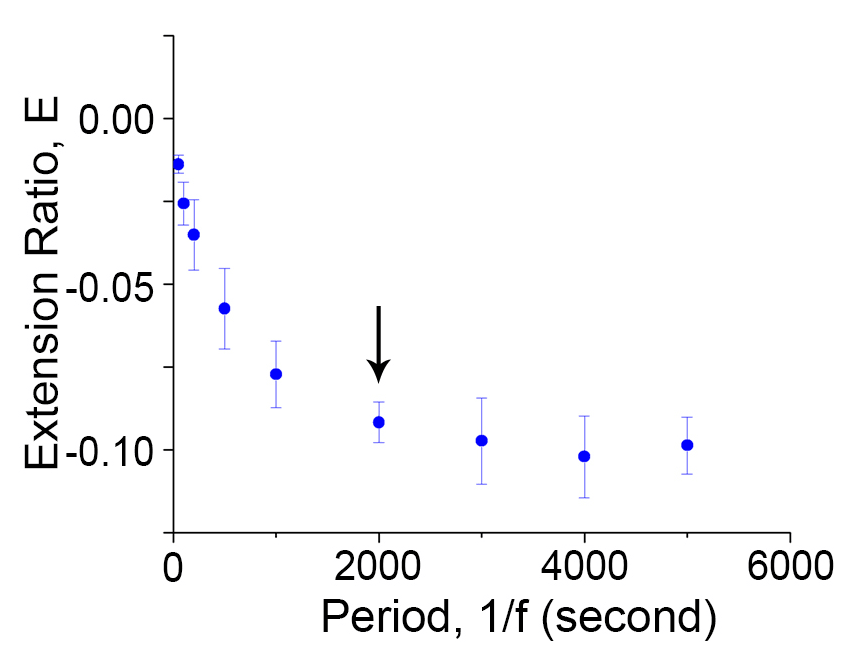} \\
	
    \end{tabular}
    
    \caption{Magnetostriction of binary colloidal crystals. Experiments were performed in a magnetic field with a constant vertical component of 9.5 Oe and a sinusoidally varying horizontal field with a magnitude of 2.1 Oe and a frequency of $5\times10^{-4}$ Hz.  (a) General deformation of a rectangle. The black rectangle with lengths $a_{\mathrm{i}}$ and $b_{\mathrm{i}}$ represents the unstressed state, and the blue shape represents the deformed state with lengths, $a_{\mathrm{f}}$, $b_{\mathrm{f}}$, $c$, and $d$.  An example experimental region used for measuring the magnetostriction effect is illustrated in both (b) real-space and (c) reciprocal-space. The Joule magnetostriction coefficient, defined as the extension ratio E is plotted in (d) as a function of the tilt angle for fixed crystal orientation $\psi = 0^{\circ}$, and (e) as a function of the crystal orientation for a fixed tilt angle $\theta = 12.5^{\circ}$.  The insets in (d) illustrate an exaggerated crystal compression for small and large tilt angles for a fixed $\psi = 0^{\circ}$. The insets in (e) illustrate the field direction relative to the crystal orientation for $\psi = 2.5^{\circ}$ and $\psi = 32.5^{\circ}$, respectively, with fixed $\theta = 12.5^{\circ}$. The extension ratio as a function of tilting frequency is illustrated in (f), where $1/f$ corresponds to the period of the in-plane sinusoidal wave. The black arrow indicates the tilting frequency that is used for obtaining the data in panels (d) and (e).}
    \label{fig:Figure6}
\end{figure*}

Tilting the magnetic field away from the vertical induces magnetostrictive compression for tilt angles $\theta > 0^{\circ}$. Experimentally, a strong vertical field ($9.5\,$Oe) is applied, in order to reduce thermal motion, and a weak in-plane sinusoidal magnetic field is supplemented.  The in-plane component was set to oscillate at $5.00\times10^{-4}$ Hz, so that the crystal had sufficient time to equilibrate. Repeating experiments at rates corresponding to oscillation periods between $2.00\times 10^{3}$s to $5.00\times 10^{3}$s gave comparable results (Fig.~\ref{fig:Figure6}f), suggesting that our experiments were performed at (or near) thermodynamic equilibrium.

Figure~\ref{fig:Figure6} illustrates the magnetostriction coefficient for different $\theta$ and crystal orientations, $\psi$, relative to the in-plane component of the field. The components of the 2D strain tensor are given by $\varepsilon_{xx} = (a_{\mathrm{f}} - a_{\mathrm{i}})/a_{\mathrm{i}}$, $\varepsilon_{yy} = (b_{\mathrm{f}} - b_{\mathrm{i}})/b_{\mathrm{i}}$, $\varepsilon_{xy} = c/a_{\mathrm{i}}$, and  $\varepsilon_{yx} = d/b_{\mathrm{i}}$, where $a_{\mathrm{j}}$ and $b_{\mathrm{j}}$ are the unstressed ($\theta = 0^{\circ}$) lattice constants in the $x$ and $y$ directions, respectively, and $a_{\mathrm{f}}$, $b_{\mathrm{f}}$, $c$, and $d$ are measured in the final state (Fig.~\ref{fig:Figure6}a), using the original lattice in the experimental image.  The non-magnetic particles at the corners of the rectangle depicted in Fig.~\ref{fig:Figure6}b were used for the measurements. Both $a_{\mathrm{i}}$ and $b_{\mathrm{i}}$ are approximately equal to $\sqrt{2}(\sigma_{\mathrm{m}} + \sigma_{\mathrm{n}})/2$. We find that dilation $(\varepsilon_{xx} + \varepsilon_{yy})$, rotation $(\varepsilon_{xy} - \varepsilon_{yx})$, and transverse shear $(\varepsilon_{xy} + \varepsilon_{yx})$ are all negligible. Figure~\ref{fig:Figure6} thus only depicts the extension ratio (or longitudinal shear) $\mathrm{E} = \varepsilon_{xx} + \varepsilon_{yy}$, which results from Joule magnetostriction.  Negative values indicate that the crystal is compressed along the direction of the in-plane field. For $\psi = 0^{\circ}$ (a field aligned with the 10 crystal direction) the magnetostriction is maximal, exceeding 10\%, which is an order of magnitude larger than that observed in giant magnetostrictive materials;\cite{kainuma} by contrast, for $\psi = 45^{\circ}$ (a field along the 11 crystal direction), magnetostriction is suppressed (see Movie S3). To test the accuracy of our measurement of Joule magnetostriction, the reciprocal lattice obtained via Fast Fourier Transform (FFT) of the real-space images was also analyzed. The peaks at the corners of the rectangle in Fig.~\ref{fig:Figure6}c were used to calculate the magnetostriction coefficient. The difference between the real- and reciprocal-space results is within experimental uncertainty. 

\subsection{Large tilt angles: martensitic transformations}
\label{sec:martensitic}

\begin{figure*}
    \centering
    
       \begin{tabular}{r c}
	&\multicolumn{1}{l}{\hspace{0mm} $~~~~~~~~~~~~\theta = 0^{\circ}$ \hspace{19mm} $\theta = 16^{\circ}$ \hspace{19mm} $\theta = 21^{\circ}$ \hspace{19mm} $\theta = 50^{\circ}$}\\
	\begin{minipage}[b]{2.31cm}{ (a)  \\ \hspace*{4mm} Simulation\\\hspace*{4mm} $H = 12\,\mathrm{Oe}$\\ \hspace*{4mm} $\Psi = 0^{\circ}$\\  \hspace*{4mm} 2D confined \\ \\ } \end{minipage}& \includegraphics[width=0.7\textwidth]{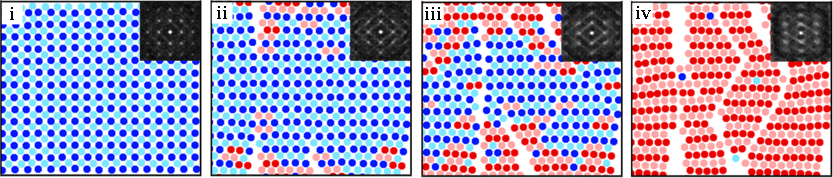}\\
	\begin{minipage}[b]{2.31cm}{ (b)  \\ \hspace*{4mm} Experiment\\\hspace*{4mm} $H = 12\,\mathrm{Oe}$\\ \hspace*{4mm} $\Psi = 0^{\circ}$\\  \\ \\} \end{minipage}&\includegraphics[width=0.7\textwidth]{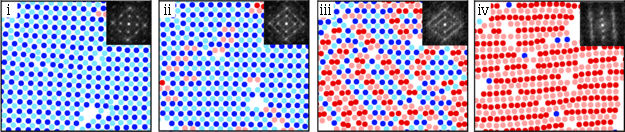}\\
	\begin{minipage}[b]{2.31cm}{ (c)  \\ \hspace*{4mm} Experiment \\\hspace*{4mm} $H = 5\,\mathrm{Oe}$\\ \hspace*{4mm} $\Psi = 0^{\circ}$\\  \\ \\} \end{minipage}&\includegraphics[width=0.7\textwidth]{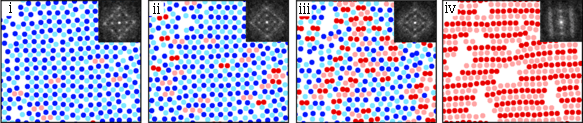}\\
	\begin{minipage}[b]{2.31cm}{ (d)  \\ \hspace*{4mm} Simulation\\\hspace*{4mm} $H = 12\,\mathrm{Oe}$\\ \hspace*{4mm} $\Psi = 45^{\circ}$\\  \hspace*{4mm} 2D confined \\ \\} \end{minipage}&\includegraphics[width=0.7\textwidth]{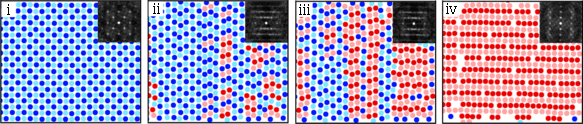}\\  
	\begin{minipage}[b]{2.31cm}{ (e)  \\ \hspace*{4mm} Experiment\\\hspace*{4mm} $H = 12\,\mathrm{Oe}$\\ \hspace*{4mm} $\Psi = 45^{\circ}$\\  \\ \\} \end{minipage}&\includegraphics[width=0.7\textwidth]{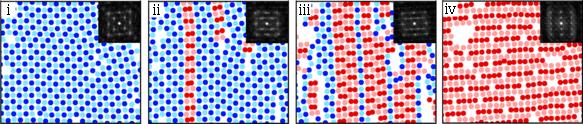}\\  
	\begin{minipage}[b]{2.31cm}{ (f)  \\ \hspace*{4mm} Experiment\\\hspace*{4mm} $H = 5\,\mathrm{Oe}$\\ \hspace*{4mm} $\Psi = 45^{\circ}$\\  \\ \\} \end{minipage}&\includegraphics[width=0.7\textwidth]{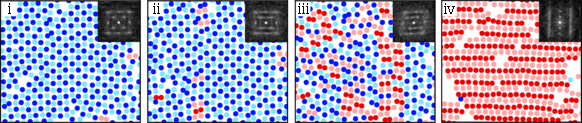}\\
	
    \end{tabular}

    \caption{False color experimental and simulation images of the martensitic transformation from a checkerboard crystal phase to a striped crystal phase for different crystal orientations and field strengths. Each row shows a sequence of images from a single experiment or simulation in which the tilt angle of the field was monotonically increased. For visualization purposes, the particles with at least one martensitic bond are falsely colored red (pink) if they are magnetic (nonmagnetic), while particles with only austenitic bonds are colored dark blue (light blue) if they are magnetic (nonmagnetic). Simulated transformations under perfect 2D confinement and experimental transformations at both strong (12 Oe) and weak (5 Oe) fields are shown for each of the two crystal orientations $\psi = 0^{\circ}$ and $\psi = 45^{\circ}$. By setting $h=1.11\sigma_{n}$ in simulations (Movies S9 and S11), better agreement with experiment is obtained, which highlights the role of confinement in the transformation dynamics. In all experiments, the tilt rate of the field $0.008^{\circ}/$s.} 
    \label{fig:Figure7}
\end{figure*}

To explore the transformation from checkerboard to striped crystal, we applied a rotating magnetic field with constant angular frequency and field strength. For $\theta > 15^{\circ}$, the crystal undergoes a diffusionless transformation through a combination of compression, shear, and slip along the 10 or 11 lattice directions. Video microscopy and numerical simulations reveal a variety of possible combinations of these different elements (Fig.~\ref{fig:Figure7}). For a fixed $T$ (constant $H$), the chosen pathway indeed depends on the initial crystal orientation, field strength, and degree of planar confinement. Note that the latter two factors are related because out-of-plane motion is suppressed near the melting transition, where the gravitational cost of buckling is large compared with the energy gain of forming dimers.

To characterize the different experimental transformation pathways, we first consider the behavior of systems where the motion is perfectly confined to a 2D plane, which are only accessible in simulation. In this case, for all field strengths the transition at $\psi = 0 ^{\circ}$ proceeds through continuous longitudinal shear (see Fig.~\ref{fig:Figure7}(row a) and Movie S4 for the high field cases, and Movie S5 for the low field cases), while at $\psi = 45^{\circ}$ zigzags form and gradually straighten into horizontal stripes (see Fig.~\ref{fig:Figure7}(row d) and Movie S6 for the high field cases, and Movie S7 for the low field cases). These pathways are qualitatively different from those observed in experiments (see Fig.~\ref{fig:Figure7}(rows b, c, e, and f)), in which perfect 2D confinement is not accessible.  Although we cannot precisely quantify the amount of experimental buckling, we can estimate it by comparing the distance between adjacent particle centers relative to the diameter of the particle. This analysis suggests that in most experiments the buckling angle between two adjacent particles was in the range of $5^{\circ}$ to $15^{\circ}$ relative to the plane.   When a similar amount of out-of-plane motion (buckling) is allowed in simulation, we find markedly improved agreement with experiment and thus conclude that confinement plays a key role in tuning the transformation pathway. 

Each row in Fig.~\ref{fig:Figure7} shows the crystal transformations for different tilt angles $\theta$ and crystal orientations $\psi$.  The first and fourth rows depict the simulated transformation for strict 2D confinement. The second and fifth rows depict the experimental transformation in a strong field. The third and sixth rows depict the corresponding experimental transformations in a weak field.

Experimental observations of the transition from checkerboard to striped phase for two field strengths and two initial crystal orientations are shown in Fig.~\ref{fig:Figure7}.  The experimentally observed transformation pathway for a strong field and $\psi \approx 0^{\circ}$ proceeds first through magnetostrictive compression, and then by shear, resulting in the diagonal lines of dimers visible in Fig.~\ref{fig:Figure7}b(iii).  Adjacent dimer lines then align and coalesce to form the striped phase of Fig.~\ref{fig:Figure7}b(iv).  (See experimental Movie S8 and simulation Movie S9, which allows buckling, unlike the simulation of Fig.~\ref{fig:Figure7}a(iii)). These transformations are smooth and homogeneous. By contrast, the transformation for $\psi \approx 45^{\circ}$ proceeds through a sequence of abrupt slips, giving rise to the vertical lines of nearly horizontal dimers visible in Figs.~\ref{fig:Figure7}e(ii) and~\ref{fig:Figure7}e(iii).  In most cases, the dimer line nucleates at a defect site and is facilitated by a local lattice expansion, which opens up free space for adjacent columns to shift and for dimer lines to zip up.  Isolated dimer lines first form throughout the crystal, typically appearing every few columns.  Dimer lines and monomer lines then coalesce into trimers or quadrimers (see experimental Movie S10 and simulation Movie S11), depending on the spacing between nearby dimer lines, and finally form horizontal stripes (Fig.~\ref{fig:Figure7}e(iv)). The sudden coordinated line slips result in structures that resemble martensitic plates.\cite{liu, waitz} For weaker fields, the features observed at strong fields remain qualitatively discernible, but the coherence of the dimer lines is weakened, as displayed in Figs.~\ref{fig:Figure7}c(iii) and~\ref{fig:Figure7}f(iii) (see experimental Movies S12 and S13, and simulation Movies S14 and S15). We attribute this effect to the increased importance of gravity relative to the dipole interactions, which suppresses buckling and thus hybridizes the pathways with those of perfectly confined systems.  The perfectly 2D confined systems, which can only be probed via simulations, show little evidence of a dimer phase, as may be expected given that dimer formation in the experimental system involves buckling (see Figs.~\ref{fig:Figure7}a(iii) and ~\ref{fig:Figure7}d(iii)).

Fourier transforms of the experimental images (see insets in Fig.~\ref{fig:Figure7} and Movies S8, S10, S12, and S13) confirm the rotational symmetries of the different intermediates and show evidence of ordering at wavenumbers present in neither the initial checkerboard not the final striped phase. For $\psi \approx 0^{\circ}$, the system is first continuously sheared.  Let $\vec{q}_1=(q_1,0)$ and $\vec{q}_2=(0,q_2)$ be the wave vectors of the fundamental Fourier peaks.  During the magnetostriction, $q_1$ grows and $q_2$ shrinks.  As dimer lines form, diffuse Fourier peaks appear at $\pm(-\vec{q}_1 + \vec{q}_2)/2$ (Fig.~\ref{fig:Figure7}b(iii) and Movie S8), indicating the emergence of a dimer configuration with a unit cell twice as large as the (sheared) checkerboard.  As the field is tilted further, however, both the peaks at $(-\vec{q}_1\pm\vec{q}_2)/2$ and at $\vec{q}_1$ are extinguished, a signature of the striped phase.  While trimer or quadrimer patterns may also exist, we see no clear evidence of ordering at the wavenumbers associated with them.

For $\psi \approx 45^{\circ}$, the rectangular symmetry of the Fourier pattern is retained throughout the transformation, and Fourier peaks remain fixed in position, indicating that there is no overall shearing or compression of the lattice (see Movie S10).  Let $\vec{q}_3=(1,1)q$ and $\vec{q}_4=(1,-1)q$ be the wave vectors of the fundamental Fourier peaks in the checkerboard phase in this orientation. The presence of an ordered dimer configuration is then signaled by the emergence of Fourier peaks at $(\vec{q}_3+\vec{q}_4)/4 \pm (\vec{q}_3-\vec{q}_4)/2 = (1/2,\pm1)q$ (see Fig.~\ref{fig:Figure7}e(iii) and Movie S10),  which then vanish as the dimers join to form longer chains and eventually extended stripes.

Fourier images of the striped phase display two columns of peaks at $q_x \approx \pm 2q$.  The peaks at nonzero values of $q_x$ show no evidence of different spacings for the two stripe types.  Though potential energy calculations indicate that an incommensurate phase is the ground state for a tilt angle of $50^{\circ}$, the Fourier images indicate that at finite temperatures the system does not display this feature.

\section{Conclusions}
\label{sec:conclusions}

We have observed several different types of fluid-solid and solid-solid transformations in a binary system of magnetic and nonmagnetic spheres. By exploring the mechanism behind fluid-solid transformation, we were able to optimize the experimental conditions so that a large single domain checkerboard lattice can grow within a few hours. By probing the solid-solid transformations with a tilted magnetic field, we also found that the diffusionless transformation from a checkerboard to a striped phase can be selected by choosing the applied field strength and the orientation of the initial crystal with respect to the tilt direction of the applied field.  For high fields, the different pathways both pass through intermediate states dominated by buckled dimers before reaching the striped phase.  Because of the slow relaxation timescales along the transformation pathway, we suspect that the experiment is not performed in the quasistatic regime. Dimer formation and alignment may thus be a far from equilibrium feature. We note, however, that potential energy calculations suggest that at least two different dimer phases may be stable at certain tilt angles (see Fig.~\ref{fig:Figure2}).  At low temperatures (or high field strengths), phases of trimers, quadrimers, \emph{etc.} may also play a role if the field can be tilted sufficiently slowly.  For the relatively rapid tilt rates employed in our experiments, the high degree of order observed quite far from equilibrium is rather remarkable.

In general, the ability to switch among different pathways could enable the control of functional characteristics of engineered materials, including their capacity for heat exchange and susceptibility to shape change along the transformation.  The binary magnetic system described here also opens the way for studying other dynamical processes in alloys. For instance, changing the ferrofluid susceptibility would enable the investigation of solid-solid transitions that require long-range diffusion, such as the transformation from a striped crystal to a hexagonal crystal\cite{khalil} with $n_{\mathrm{m}}/n_{\mathrm{n}}=2$ (see Fig.~\ref{fig:Figure1}d).

\section*{Acknowledgements}

The authors are thankful for support from the National Science Foundation Research Triangle Materials Research Science and Engineering Center (DMR-1121107), the China Youth 1000 Scholars program, and the National Science Foundation of China (NSFC-51350110334).

\footnotesize{
\providecommand*{\mcitethebibliography}{\thebibliography}
\csname @ifundefined\endcsname{endmcitethebibliography}
{\let\endmcitethebibliography\endthebibliography}{}

}


\begin{mcitethebibliography}{37}
\providecommand*{\natexlab}[1]{#1}
\providecommand*{\mciteSetBstSublistMode}[1]{}
\providecommand*{\mciteSetBstMaxWidthForm}[2]{}
\providecommand*{\mciteBstWouldAddEndPuncttrue}
  {\def\EndOfBibitem{\unskip.}}
\providecommand*{\mciteBstWouldAddEndPunctfalse}
  {\let\EndOfBibitem\relax}
\providecommand*{\mciteSetBstMidEndSepPunct}[3]{}
\providecommand*{\mciteSetBstSublistLabelBeginEnd}[3]{}
\providecommand*{\EndOfBibitem}{}
\mciteSetBstSublistMode{f}
\mciteSetBstMaxWidthForm{subitem}
{(\emph{\alph{mcitesubitemcount}})}
\mciteSetBstSublistLabelBeginEnd{\mcitemaxwidthsubitemform\space}
{\relax}{\relax}

\bibitem[Kainuma \emph{et~al.}(2006)Kainuma, Imano, Ito, Sutou, Morito,
  Okamoto, Kitakami, Oikawa, Fujita, Kanomata, and Ishida]{kainuma}
R.~Kainuma, Y.~Imano, W.~Ito, Y.~Sutou, H.~Morito, S.~Okamoto, O.~Kitakami,
  K.~Oikawa, A.~Fujita, T.~Kanomata and K.~Ishida, \emph{Nature}, 2006,
  \textbf{439}, 957--960\relax
\mciteBstWouldAddEndPuncttrue
\mciteSetBstMidEndSepPunct{\mcitedefaultmidpunct}
{\mcitedefaultendpunct}{\mcitedefaultseppunct}\relax
\EndOfBibitem
\bibitem[Tanaka \emph{et~al.}(2010)Tanaka, Himuro, Kainuma, Sutou, Omori, and
  Ishida]{tanaka}
Y.~Tanaka, Y.~Himuro, R.~Kainuma, Y.~Sutou, T.~Omori and K.~Ishida,
  \emph{Science}, 2010, \textbf{327}, 1488--1490\relax
\mciteBstWouldAddEndPuncttrue
\mciteSetBstMidEndSepPunct{\mcitedefaultmidpunct}
{\mcitedefaultendpunct}{\mcitedefaultseppunct}\relax
\EndOfBibitem
\bibitem[Angel(1954)]{angel}
T.~Angel, \emph{Journal of the Iron and Steel Institute}, 1954, \textbf{177},
  165--174\relax
\mciteBstWouldAddEndPuncttrue
\mciteSetBstMidEndSepPunct{\mcitedefaultmidpunct}
{\mcitedefaultendpunct}{\mcitedefaultseppunct}\relax
\EndOfBibitem
\bibitem[Moya \emph{et~al.}(2012)Moya, Hueso, Maccherozzi, Tovstolytkin,
  Podyalovskii, Ducati, Phillips, Ghidini, Hovorka, Berger, Defay, Dhesi, and
  Mathur]{moya}
X.~Moya, L.~E. Hueso, F.~Maccherozzi, A.~I. Tovstolytkin, D.~I. Podyalovskii,
  C.~Ducati, L.~C. Phillips, M.~Ghidini, O.~Hovorka, M.~E. Berger, E.~Defay,
  S.~S. Dhesi and N.~D. Mathur, \emph{Nature Materials}, 2012, \textbf{12},
  52--58\relax
\mciteBstWouldAddEndPuncttrue
\mciteSetBstMidEndSepPunct{\mcitedefaultmidpunct}
{\mcitedefaultendpunct}{\mcitedefaultseppunct}\relax
\EndOfBibitem
\bibitem[Liu \emph{et~al.}(2011)Liu, Gottschall, Skokov, Moore, and
  Gutfleish]{liu}
J.~Liu, T.~Gottschall, K.~P. Skokov, J.~D. Moore and O.~Gutfleish, \emph{Nature
  Materials}, 2011, \textbf{11}, 620--626\relax
\mciteBstWouldAddEndPuncttrue
\mciteSetBstMidEndSepPunct{\mcitedefaultmidpunct}
{\mcitedefaultendpunct}{\mcitedefaultseppunct}\relax
\EndOfBibitem
\bibitem[Song \emph{et~al.}(2011)Song, Chen, Dabade, Shield, and James]{song}
Y.~Song, X.~Chen, V.~Dabade, T.~W. Shield and R.~D. James, \emph{Nature}, 2011,
  \textbf{502}, 85--88\relax
\mciteBstWouldAddEndPuncttrue
\mciteSetBstMidEndSepPunct{\mcitedefaultmidpunct}
{\mcitedefaultendpunct}{\mcitedefaultseppunct}\relax
\EndOfBibitem
\bibitem[Bhattacharya \emph{et~al.}(2004)Bhattacharya, Conti, Zanzotto, and
  Zimmer]{bhattacharya}
K.~Bhattacharya, S.~Conti, G.~Zanzotto and J.~Zimmer, \emph{Nature}, 2004,
  \textbf{428}, 55--59\relax
\mciteBstWouldAddEndPuncttrue
\mciteSetBstMidEndSepPunct{\mcitedefaultmidpunct}
{\mcitedefaultendpunct}{\mcitedefaultseppunct}\relax
\EndOfBibitem
\bibitem[Chernenko \emph{et~al.}(1998)Chernenko, Segu\'\i, Cesari, Pons, and
  Kokorin]{chernenko}
V.~A. Chernenko, C.~Segu\'\i, E.~Cesari, J.~Pons and V.~V. Kokorin,
  \emph{Physical Review B}, 1998, \textbf{57}, 2659--2662\relax
\mciteBstWouldAddEndPuncttrue
\mciteSetBstMidEndSepPunct{\mcitedefaultmidpunct}
{\mcitedefaultendpunct}{\mcitedefaultseppunct}\relax
\EndOfBibitem
\bibitem[Meyers \emph{et~al.}(2002)Meyers, Perez-Prado, Xue, Xu, and
  McNelley]{meyers}
M.~A. Meyers, M.~T. Perez-Prado, Q.~Xue, Y.~Xu and T.~R. McNelley, \emph{AIP
  Conference Proceedings}, 2002, \textbf{620}, 571--574\relax
\mciteBstWouldAddEndPuncttrue
\mciteSetBstMidEndSepPunct{\mcitedefaultmidpunct}
{\mcitedefaultendpunct}{\mcitedefaultseppunct}\relax
\EndOfBibitem
\bibitem[Waitz \emph{et~al.}(2004)Waitz, Kazykhanov, and Karnthaler]{waitz}
T.~Waitz, V.~Kazykhanov and H.~Karnthaler, \emph{Acta Materialia}, 2004,
  \textbf{52}, 137 -- 147\relax
\mciteBstWouldAddEndPuncttrue
\mciteSetBstMidEndSepPunct{\mcitedefaultmidpunct}
{\mcitedefaultendpunct}{\mcitedefaultseppunct}\relax
\EndOfBibitem
\bibitem[Kadau \emph{et~al.}(2002)Kadau, Germann, Lomdahl, and Holian]{kadau}
K.~Kadau, T.~C. Germann, P.~S. Lomdahl and B.~L. Holian, \emph{Science}, 2002,
  \textbf{296}, 1681--1684\relax
\mciteBstWouldAddEndPuncttrue
\mciteSetBstMidEndSepPunct{\mcitedefaultmidpunct}
{\mcitedefaultendpunct}{\mcitedefaultseppunct}\relax
\EndOfBibitem
\bibitem[Gasser \emph{et~al.}(2001)Gasser, Weeks, Schofield, Pusey, and
  Weitz]{gasser}
U.~Gasser, E.~R. Weeks, A.~Schofield, P.~N. Pusey and D.~A. Weitz,
  \emph{Science}, 2001, \textbf{292}, 258--262\relax
\mciteBstWouldAddEndPuncttrue
\mciteSetBstMidEndSepPunct{\mcitedefaultmidpunct}
{\mcitedefaultendpunct}{\mcitedefaultseppunct}\relax
\EndOfBibitem
\bibitem[Leunissen \emph{et~al.}(2005)Leunissen, Christova, Hynninen, Royall,
  Campbell, Imhof, Dijkstra, van Roij, and van Blaaderen]{leunissen}
M.~E. Leunissen, C.~G. Christova, A.-P. Hynninen, C.~P. Royall, A.~I. Campbell,
  A.~Imhof, M.~Dijkstra, R.~van Roij and A.~van Blaaderen, \emph{Nature}, 2005,
  \textbf{437}, 235--240\relax
\mciteBstWouldAddEndPuncttrue
\mciteSetBstMidEndSepPunct{\mcitedefaultmidpunct}
{\mcitedefaultendpunct}{\mcitedefaultseppunct}\relax
\EndOfBibitem
\bibitem[Chen \emph{et~al.}(2011)Chen, Bae, and Granick]{chen_Q}
Q.~Chen, S.~C. Bae and S.~Granick, \emph{Nature}, 2011, \textbf{469},
  381--384\relax
\mciteBstWouldAddEndPuncttrue
\mciteSetBstMidEndSepPunct{\mcitedefaultmidpunct}
{\mcitedefaultendpunct}{\mcitedefaultseppunct}\relax
\EndOfBibitem
\bibitem[Evers \emph{et~al.}(2010)Evers, Nijs, Filion, Castillo, Dijkstra, and
  Vanmaekelbergh]{evers}
W.~H. Evers, B.~D. Nijs, L.~Filion, S.~Castillo, M.~Dijkstra and
  D.~Vanmaekelbergh, \emph{Nano Letters}, 2010, \textbf{10}, 4235--4241\relax
\mciteBstWouldAddEndPuncttrue
\mciteSetBstMidEndSepPunct{\mcitedefaultmidpunct}
{\mcitedefaultendpunct}{\mcitedefaultseppunct}\relax
\EndOfBibitem
\bibitem[Talapin \emph{et~al.}(2009)Talapin, Shevchenko, Bodnarchuk, Ye, Chen,
  and Murray]{talapin}
D.~V. Talapin, E.~V. Shevchenko, M.~I. Bodnarchuk, X.~Ye, J.~Chen and C.~B.
  Murray, \emph{Nature}, 2009, \textbf{461}, 964--967\relax
\mciteBstWouldAddEndPuncttrue
\mciteSetBstMidEndSepPunct{\mcitedefaultmidpunct}
{\mcitedefaultendpunct}{\mcitedefaultseppunct}\relax
\EndOfBibitem
\bibitem[Shevchenko \emph{et~al.}(2006)Shevchenko, Talapin, Kotov, O'Brien, and
  Murray]{shevchenko}
E.~V. Shevchenko, D.~V. Talapin, N.~A. Kotov, S.~O'Brien and C.~B. Murray,
  \emph{Nature}, 2006, \textbf{439}, 55--59\relax
\mciteBstWouldAddEndPuncttrue
\mciteSetBstMidEndSepPunct{\mcitedefaultmidpunct}
{\mcitedefaultendpunct}{\mcitedefaultseppunct}\relax
\EndOfBibitem
\bibitem[Shereda \emph{et~al.}(2010)Shereda, Larson, and Solomon]{shereda}
L.~T. Shereda, R.~G. Larson and M.~J. Solomon, \emph{Physical Review Letters},
  2010, \textbf{105}, 228302\relax
\mciteBstWouldAddEndPuncttrue
\mciteSetBstMidEndSepPunct{\mcitedefaultmidpunct}
{\mcitedefaultendpunct}{\mcitedefaultseppunct}\relax
\EndOfBibitem
\bibitem[Bubeck \emph{et~al.}(1999)Bubeck, Bechinger, Neser, and
  Leiderer]{bubeck}
R.~Bubeck, C.~Bechinger, S.~Neser and P.~Leiderer, \emph{Physical Review
  Letters}, 1999, \textbf{82}, 3364--3367\relax
\mciteBstWouldAddEndPuncttrue
\mciteSetBstMidEndSepPunct{\mcitedefaultmidpunct}
{\mcitedefaultendpunct}{\mcitedefaultseppunct}\relax
\EndOfBibitem
\bibitem[Wang \emph{et~al.}(2012)Wang, Wang, Peng, Zheng, and Han]{wang}
Z.~Wang, F.~Wang, Y.~Peng, Z.~Zheng and Y.~Han, \emph{Science}, 2012,
  \textbf{338}, 87--90\relax
\mciteBstWouldAddEndPuncttrue
\mciteSetBstMidEndSepPunct{\mcitedefaultmidpunct}
{\mcitedefaultendpunct}{\mcitedefaultseppunct}\relax
\EndOfBibitem
\bibitem[Yethiraj \emph{et~al.}(2004)Yethiraj, Wouterse, Groh, and van
  Blaaderen]{yethiraj_wouterse}
A.~Yethiraj, A.~Wouterse, B.~Groh and A.~van Blaaderen, \emph{Physical Review
  Letters}, 2004, \textbf{92}, 058301\relax
\mciteBstWouldAddEndPuncttrue
\mciteSetBstMidEndSepPunct{\mcitedefaultmidpunct}
{\mcitedefaultendpunct}{\mcitedefaultseppunct}\relax
\EndOfBibitem
\bibitem[Yethiraj and van Blaaderen(2003)]{yethiraj_blaaderen}
A.~Yethiraj and A.~van Blaaderen, \emph{Nature}, 2003, \textbf{421},
  513--517\relax
\mciteBstWouldAddEndPuncttrue
\mciteSetBstMidEndSepPunct{\mcitedefaultmidpunct}
{\mcitedefaultendpunct}{\mcitedefaultseppunct}\relax
\EndOfBibitem
\bibitem[Han \emph{et~al.}(2008)Han, Shokef, Alsayed, Yunker, Lubensky, and
  Yodh]{han}
Y.~Han, Y.~Shokef, A.~M. Alsayed, P.~Yunker, T.~C. Lubensky and A.~G. Yodh,
  \emph{Nature}, 2008, \textbf{456}, 898--903\relax
\mciteBstWouldAddEndPuncttrue
\mciteSetBstMidEndSepPunct{\mcitedefaultmidpunct}
{\mcitedefaultendpunct}{\mcitedefaultseppunct}\relax
\EndOfBibitem
\bibitem[Leunissen \emph{et~al.}(2009)Leunissen, Vutukuri, and van
  Blaaderen]{leunissen_vutukuri}
M.~E. Leunissen, H.~R. Vutukuri and A.~van Blaaderen, \emph{Advanced
  Materials}, 2009, \textbf{21}, 3116--3120\relax
\mciteBstWouldAddEndPuncttrue
\mciteSetBstMidEndSepPunct{\mcitedefaultmidpunct}
{\mcitedefaultendpunct}{\mcitedefaultseppunct}\relax
\EndOfBibitem
\bibitem[Peng \emph{et~al.}(2015)Peng, Wang, Alsayed, Zhang, Yodh, and
  Han]{peng}
Y.~Peng, F.~Wang, A.~M. Alsayed, Z.~Zhang, A.~G. Yodh and Y.~Han, \emph{Nature
  Materials}, 2015, \textbf{14}, 101--108\relax
\mciteBstWouldAddEndPuncttrue
\mciteSetBstMidEndSepPunct{\mcitedefaultmidpunct}
{\mcitedefaultendpunct}{\mcitedefaultseppunct}\relax
\EndOfBibitem
\bibitem[Weiss \emph{et~al.}(1995)Weiss, Oxtoby, Grier, and Murray]{weiss}
J.~A. Weiss, D.~W. Oxtoby, D.~G. Grier and C.~A. Murray, \emph{The Journal of
  Chemical Physics}, 1995, \textbf{103}, 1180--1190\relax
\mciteBstWouldAddEndPuncttrue
\mciteSetBstMidEndSepPunct{\mcitedefaultmidpunct}
{\mcitedefaultendpunct}{\mcitedefaultseppunct}\relax
\EndOfBibitem
\bibitem[Casey \emph{et~al.}(2012)Casey, Scarlett, Rogers, Jenkins, Sinno, and
  Crocker]{casey}
M.~T. Casey, R.~T. Scarlett, B.~Rogers, I.~Jenkins, T.~Sinno and J.~C. Crocker,
  \emph{Nature Communications}, 2012, \textbf{3}, 1209\relax
\mciteBstWouldAddEndPuncttrue
\mciteSetBstMidEndSepPunct{\mcitedefaultmidpunct}
{\mcitedefaultendpunct}{\mcitedefaultseppunct}\relax
\EndOfBibitem
\bibitem[Griffiths(1999)]{griffiths}
D.~J. Griffiths, \emph{Introduction to Electrodynamics}, Prentice Hall,
  1999\relax
\mciteBstWouldAddEndPuncttrue
\mciteSetBstMidEndSepPunct{\mcitedefaultmidpunct}
{\mcitedefaultendpunct}{\mcitedefaultseppunct}\relax
\EndOfBibitem
\bibitem[Jackson(1999. See Problem 5.17.)]{jackson}
J.~D. Jackson, \emph{Classical Electrodynamics}, John Wiley \& Sons, 3rd edn,
  1999. See Problem 5.17.\relax
\mciteBstWouldAddEndPunctfalse
\mciteSetBstMidEndSepPunct{\mcitedefaultmidpunct}
{}{\mcitedefaultseppunct}\relax
\EndOfBibitem
\bibitem[Khalil \emph{et~al.}(2012)Khalil, Sagastegui, Li, Tahir, Socolar,
  Wiley, and Yellen]{khalil}
K.~S. Khalil, A.~Sagastegui, Y.~Li, M.~A. Tahir, J.~E.~S. Socolar, B.~J. Wiley
  and B.~B. Yellen, \emph{Nature Communications}, 2012, \textbf{3}, 794\relax
\mciteBstWouldAddEndPuncttrue
\mciteSetBstMidEndSepPunct{\mcitedefaultmidpunct}
{\mcitedefaultendpunct}{\mcitedefaultseppunct}\relax
\EndOfBibitem
\bibitem[Filion and Dijkstra(2009)]{filion}
L.~Filion and M.~Dijkstra, \emph{Physical Review E}, 2009, \textbf{79},
  046714\relax
\mciteBstWouldAddEndPuncttrue
\mciteSetBstMidEndSepPunct{\mcitedefaultmidpunct}
{\mcitedefaultendpunct}{\mcitedefaultseppunct}\relax
\EndOfBibitem
\bibitem[Schoen and Klapp(2007)]{lipkowitz}
M.~Schoen and S.~H.~L. Klapp, Reviews in Computational Chemistry, 2007\relax
\mciteBstWouldAddEndPuncttrue
\mciteSetBstMidEndSepPunct{\mcitedefaultmidpunct}
{\mcitedefaultendpunct}{\mcitedefaultseppunct}\relax
\EndOfBibitem
\bibitem[Chen and Siepmann(2001)]{chen_B}
B.~Chen and J.~I. Siepmann, \emph{Journal of Physical Chemistry B}, 2001,
  \textbf{105}, 11275--11282\relax
\mciteBstWouldAddEndPuncttrue
\mciteSetBstMidEndSepPunct{\mcitedefaultmidpunct}
{\mcitedefaultendpunct}{\mcitedefaultseppunct}\relax
\EndOfBibitem
\bibitem[Almarza(2009)]{almarza}
N.~G. Almarza, \emph{The Journal of Chemical Physics}, 2009, \textbf{130},
  184106\relax
\mciteBstWouldAddEndPuncttrue
\mciteSetBstMidEndSepPunct{\mcitedefaultmidpunct}
{\mcitedefaultendpunct}{\mcitedefaultseppunct}\relax
\EndOfBibitem
\bibitem[Frenkel and Smit(2001)]{frenkel}
D.~Frenkel and B.~Smit, \emph{Understanding Molecular Simulation: From
  Algorithms to Applications}, Academic Press, 2nd edn, 2001\relax
\mciteBstWouldAddEndPuncttrue
\mciteSetBstMidEndSepPunct{\mcitedefaultmidpunct}
{\mcitedefaultendpunct}{\mcitedefaultseppunct}\relax
\EndOfBibitem
\bibitem[Suzuki \emph{et~al.}(2007)Suzuki, Kun, Yukawa, and Ito]{Ito}
M.~Suzuki, F.~Kun, S.~Yukawa and N.~Ito, \emph{Physical Review E}, 2007,
  \textbf{76}, 051116\relax
\mciteBstWouldAddEndPuncttrue
\mciteSetBstMidEndSepPunct{\mcitedefaultmidpunct}
{\mcitedefaultendpunct}{\mcitedefaultseppunct}\relax
\EndOfBibitem
\bibitem[Crocker and Grier(1996)]{crocker}
J.~C. Crocker and D.~G. Grier, \emph{Journal of Colloid and Interface Science},
  1996, \textbf{179}, 298--310\relax
\mciteBstWouldAddEndPuncttrue
\mciteSetBstMidEndSepPunct{\mcitedefaultmidpunct}
{\mcitedefaultendpunct}{\mcitedefaultseppunct}\relax
\EndOfBibitem
\end{mcitethebibliography}
\end{document}